# Molecular monolayer stabilizer for multilayer 2D materials

Cong Su[1,2,‡], Zongyou Yin[3,‡,*], Qing-Bo Yan[1,4,‡], Zegao Wang[5], Hongtao Lin[6], Lei Sun[7], Wenshuo Xu[8], Tetsuya Yamada[9], Xiang Ji[3], Nobuyuki Zettsu[9], Katsuya Teshima[9], Jamie H. Warner[8], Mircea Dincă[7], Juejun Hu[6], Mingdong Dong[5], Gang Su[11], Jing Kong[2,10], & Ju Li[1,*]

[1] Department of Nuclear and Materials Science and Engineering, Massachusetts Institute of Technology, Cambridge MA 02139

[2] Research Lab of Electronics (RLE), Massachusetts Institutes of Technology, Cambridge MA 02139

[3] Research School of Chemistry, The Australian National University, Canberra, Australian Capital Territory 2601, Australia

[4] College of Materials Science and Opto-Electronic Technology, University of Chinese Academy of Sciences, Beijing, China 100049

[5] Interdisciplinary Nanoscience Center (iNano), Aarhus University, Aarhus 8000, Denmark

[6] Department of Materials Science and Engineering, Massachusetts Institute of Technology, Cambridge MA 02139

[7] Department of Chemistry, Massachusetts Institutes of Technology, Cambridge MA 02139

[8] Department of Materials, University of Oxford, OX1 3PH, UK

[9] Center for Energy and Environmental Science, Shinshu University, 4-17-1 Wakasato, Nagano, Japan 380-8553

[10] Department of Electrical Engineering and Computer Science, Massachusetts Institute of Technology, Cambridge MA 02139

[11] School of Physical Science, University of Chinese Academy of Sciences, Beijing, China 100049

‡These authors contributed equally.

*Corresponding authors: Ju Li (liju@mit.edu), Zongyou Yin (zongyou.yin@anu.edu.au)

**Abstract**

2D van der Waals materials have rich and unique functional properties, but many are susceptible to corrosion under ambient conditions. Here we show that linear alkylamines $n\text{-}C_mH_{2m+1}NH_2$, with $m$ = 4 to 11, are highly effective in protecting the optoelectronic properties of these materials such as black phosphorous (BP) and transition metal dichalcogenides (TMDs: $WS_2$, 1T'-$MoTe_2$, $WTe_2$, $WSe_2$, $TaS_2$, and $NbSe_2$). As a representative example, *n*-hexylamine ($m$ = 6) can be applied in the form of thin molecular monolayers on BP flakes with less-than-2nm thickness and can prolong BP's lifetime from a few hours to several weeks and even months in ambient environments. Characterizations combined with our theoretical analysis show that the thin monolayers selectively sift out water molecules, forming a drying layer to achieve the passivation of the protected 2D materials. The monolayer coating is also stable in air, $H_2$ annealing, and organic solvents, but can be removed by certain organic acids.

## Introduction

Passivation of materials in air and water is foundational to our civilization (*1*). When we consider the robust ultrathin passivation of 2D materials (*2*–*7*), it should be even more essential because (a) the thickness of passivation layer on 3D materials like Si, Al, Cr etc. stays 2-5 nm over very long time, which is an insignificant fraction of the remaining unreacted bulk material. However, one cannot say this for thin 2D materials with their total thickness likely comparable to the native oxide passivation layers. Thus, the atomistic details of passivation matter even more here. (b) An ultrathin, electronically insulating layer provide opportunity to engineer extremely thin vertical heterostructures, akin to the $SiO_2$/Si gate in metal-oxide-semiconductor field-effect transistors (MOSFET). For these reasons, it is becoming increasingly critical to facilely passivate layered materials such as transition metal dichalcogenides (TMDs), black phosphorous (BP), silicene, stanine (*8*–*11*) etc., which are susceptible to corrosion under ambient conditions with air, water, or even small amounts of acidic or basic contaminants (*9*, *10*, *12*–*18*).

Several passivation strategies have been developed for these layered materials including covering by more robust 2D materials such as graphene (*19*) and hexagonal boron nitride (hBN) (*20*). However, many previous strategies suffer from processability issues and other drawbacks: Metal oxide coatings are prone to cracking (*13*, *21*); polymers (e.g. poly(methyl methacrylate) (PMMA), polystyrene (PS), Parylene, and perylene-3,4,9,10-tetracarboxylic dianhydride (PTCDA)) are readily attacked by organic solvents and offer limited durability (*18*, *22*–*25*); self-assembled monolayers with silane-terminated octadecyltrichlorosilane (OTS) are highly toxic (*26*). Here, we discovered a one-pot scalable process for passivating a large variety of 2D van der Waals materials. It involves coating a nanometer-thick monolayer of linear alkylamines onto the surface of 2D materials which greatly increases the lifetime of these materials in ambient environments with moisture and can sustain even harsh aqueous and thermal conditions. First-principles simulations suggest that the alkylamine coating significantly slows down the permeation of $O_2$, which reacts with the 2D layered material to form an ultra-thin oxide passivation layer, and completely blocks $H_2O$ molecules and shuts down the cycles of oxidation-dissolution, leading to the extended lifetime for many different classes of 2D crystals.

Since BP is the most vulnerable to corrosion among the 2D van der Waals materials studied in this work and creates the most challenges for processing, it is used here as an illustrative example of the alkylamine coating. As a representative example of linear alkylamines $n$-$C_mH_{2m+1}NH_2$, $n$-hexylamine ($m$=6) coating onto BP is systematically investigated both theoretically and experimentally in its corrosion inhibition mechanism and behaviors.

## Results

As presented in Figs. 1A-C, the coating process is divided into two steps: i) The sample together with silicon substrate is put in the liquid *n*-hexylamine for 20 minutes under 130 $^{o}$C. This step creates coating on sample, but minor cracks might exist. ii) To fix the cracks, the sample is then immersed in hexylamine vapor for another 20 minutes at 130 $^{o}$C and then annealed in argon for 30 minutes under 200 $^{o}$C after the surface cleaned by hexane. The hexane cannot remove the hexylamine coating but only the surface contamination, as shown in the later section. More detailed coating procedures are presented in the Supplementary Materials (SM) (*27*, *28*, *37–39*, *29–36*). The optimization of coating parameters of *n*-hexylamine onto BP is shown in Table S1.

Once mechanically exfoliated, the bare BP flakes are highly reactive and chemically unstable. After keeping a 3-nanometer-thick BP flake (Fig. 1D) in ambient air (humidity ~35%) for 2 days (the thickness is estimated using method from ref. (*9*)), only vague traces remain (Fig. 1E), even when care is taken to prevent light exposure, known to accelerate the damage. As shown in Fig. 1F, the three characteristic Raman peaks of BP at 361 $cm^{-1}$ ($A_g^1$), 438 $cm^{-1}$ ($B_{2g}$), and 466 $cm^{-1}$ ($A_g^2$) completely disappear after 2 days. The degradation of BP was further expedited when exposed to light, in line with previous reports (*9*) which showed that the lifetime of BP (defined as the time needed for the Raman intensity to drop to $e^{-1}$ of its original) is $\tau \approx 1$ hour when a 2.8 nm-thick sample is exposed to a photon flux of $1.8 \times 10^3$ W/cm$^2$, and $\tau \approx 10$ minutes when exposed to a photon flux of $1.7 \times 10^4$ W/cm$^2$.

In contrast, *n*-hexylamine protected BP (HA-BP hereafter), which is kept side-by-side with the unprotected one, exhibits robust BP characteristics for a much-extended period. The difference in optical contrast for HA-BP between 0 day and 111 days is essentially indiscernible (Figs. 1G and H); 31% of the intensity of $A_g^2$ was retained after 111 days (Fig. 1I). The photon fluence seen by HA-BP during Raman measurements in 111 days is equivalent to light exposure of $1.0 \times 10^5$ W/cm$^2$ for ~2 hours in total. Since such photon exposure is already known to be substantial to cause the degradation of bare BP (ref. (*9*)), we conclude that the lifetime of HA-BP can be extended even further if the sample was not exposed to the laser beam of the Raman characterization.

The coating process involves the proton transfer of the hydroxylated BP to the $-NH_2$ group of *n*-hexylamine based on the evidences below. First-principles simulations suggest that *n*-hexylamine forms a molecular monolayer as shown in Fig. 2A. The top layer of the BP surface is rapidly oxidized from the oxygen dissolved in liquid hexylamine, forming P–OH, P–O$^-$, or P=O surface groups. Experimental evidences supports a model where the acidic P–OH groups on the BP surface and the terminal $-NH_2$ groups of alkylamines undergo a Brønsted-Lowry acid-base reaction to form a layer of alkylammonium salts that coat the BP surface through a strong electrostatic interaction with the deprotonated P–O$^-$ surface sites. Confirmation that the neutral –$NH_2$ group in *n*-hexylamine becomes charged (i.e. –$NH_3^+$) came from X-ray photoelectron spectroscopy (XPS): comparing the N 1s peaks between HA-BP, dodecylamine ($C_{12}H_{25}NH_2$, R–$NH_2$), and methylammonium chloride ($CH_3NH_3Cl$, R–$NH_3^+$) reveals that HA-BP and R–$NH_3^+$ have the same binding energy, which is blue-shifted by 2.4 eV from that of R–$NH_2$ (Fig. 2B). Contact angle measurements also show that the surface of BP becomes more hydrophobic after HA coating (fig. S1), confirming that the HA coating is indeed terminated by alkyl chains, not by amine/ammonium groups.

Inspection by atomic force microscopy (AFM) of the height profile of the same 2D flake before and after coating revealed that the *n*-hexylamine coating is around 1.5 nm thick (Fig. 2C), which is consistent with the theoretical chain length of *n*-hexylamine.(*40*) This demonstrates that the deposition of *n*-hexylamine molecules is self-limiting. Polar organic solvents including acetone, ethanol, or isopropanol, as well as non-polar solvents like hexane, cannot remove the *n*-hexylamine coating, indicating that the interaction between *n*-hexylamine and BP is strong enough to sustain solvent attack. We also note that *n*-hexane does not impart any corrosion protection, attesting that the amine group is key for this function and that the alkyl chain itself cannot bind strongly on BP.

We employed first-principles calculations to investigate the transfer of protons when *n*-hexylamine approaches P–OH (Fig. 2D), formed by reacting with the water from the *n*-hexylamine coating solution. Among various structural possibilities after systematic study with results shown in figs. S2-S5, the most likely reaction pathway agrees with the scenario (P-O$^-$–$NH_3^+$-$C_6H_{13}$) proposed above and yields a bonding energy of 0.97 eV, which is 3-4 times stronger than the pure vdW interaction (~0.33 eV between *n*-hexylamine and pure BP, ~0.22 eV between amines and graphene (*41*)). The electronic density distribution shows that the H atom shares its orbital much more with N atom than with O atom (inset of Fig. 2D), and a Bader's charge analysis indicates that *n*-hexylammonium ($C_6H_{13}NH_3^+$) carries a net charge of +0.89*e*, and to compensate, the rest has -0.89*e*.

In Fig. 2E, the migration energy barrier of $H_2O$ penetrating through *n*-hexylamine is calculated to be 1.4 eV and $O_2$ 1.0 eV, when *n*-hexylamine covers BP in the densest possible packing structure (hereafter defined as 100% coverage, shown in figs. S6 and S7); when the coverage drops to 66.7%, the migration energy barrier reduces to 0.2 eV for $H_2O$ permeation and no barrier (0 eV) for $O_2$. When the HA coverage further decreases to 50% or 25%, the migration of both $H_2O$ and $O_2$ through the HA layer towards the surface of BP is barrierless. Combining this theoretical analysis with the time-evolution XPS data on phosphorous oxide concentration (Figs. 2F and G), where the oxidization speed of phosphorous after *n*-hexylamine coating is significantly reduced by 32 times at the beginning of oxidation (fitting method and definition of time constant can be found in SM), we deduce the coverage density of *n*-hexylamine on BP must be more than the defined 66.7% coverage on the surface of BP.

With these conclusions, a schematic illustration of the molecular monolayer can be shown in Fig. 2H. The top oxidized BP layer of $PO_x$ together with the coated *n*-hexylamine monolayer forms a dense protection layer for the BP underneath. It lowers down the penetration speed of $O_2$ molecule significantly and blocks the $H_2O$ molecule almost completely under room temperature, thus stabilizes the surface passivation layer (the oxidized BP at the top).

The anti-corrosion effect conferred by organic monolayer is not limited to *n*-hexylamine. Indeed, other linear alkylamines *n*-$C_mH_{2m+1}NH_2$ with $m$ = 4 to 11, including *n*-butylamine (*n*-$C_4H_9NH_2$), *n*-pentylamine (*n*-$C_5H_{11}NH_2$), *n*-octylamine (*n*-$C_8H_{17}NH_2$), *n*-decylamine (*n*- $C_{10}H_{21}NH_2$), and *n*-undecylamine (*n*-$C_{11}H_{23}NH_2$), all consistently displayed similar anti-corrosion effects in ambient air. Their coatings onto BP for anti-corrosion demonstration are presented in Table S2 in SM, and the growth parameters for coating all these alkylamines with different carbon chain lengths are summarized in Table S3.

To demonstrate the passivation efficacy for actual optoelectronic devices in ambient and aggressive environments, we fabricated two BP-flakes-based photodetectors. As a direct bandgap semiconductor, with its $E_{gap}$ continuously tunable from ~2 eV (single layer) to 0.3 eV (bulk) (*42*) by varying the number of layers, BP stands out as a promising material for photonic devices from near-infrared to mid-infrared. The layout of the uncoated BP detector with a channel length and width of ~3 μm and ~5 μm, respectively, between the Ti/Au electrodes is shown in Fig. 3A. The thickness of the BP here is 74 nm (fig. S8). The *n*-hexylamine-coated BP photodetector is shown in Fig. 3D, with comparable channel dimension and a BP thickness of 55 nm (fig. S8). The photocurrent as a function of input optical power under zero voltage bias (Figs. 3C and F, uncoated and coated respectively) was measured in ambient air with a 1550-nm laser. Both devices exhibited increased photocurrent with input power before etching (black lines labeled with pre-etching in these plots). After dipping the devices in $H_2O_2$ for 5 seconds and drying them subsequently, obvious degradation was observed under optical microscope on the uncoated BP device (Fig. 3B), while little change was found on the coated one (Fig. 3E). As evidenced by the photoelectric signal, corrosion caused severe damage to the uncoated optoelectronic device, with the photocurrent dropping to zero. In contrast, the *n*-hexylamine coated photodetector device maintained 78.6% of its original photocurrent based on the photocurrent values of 28.7 μA@post-etching and 36.5 μA@pre-etching under photoexcitation with the same input power of 3 mW. The slight drop of performance likely originates from defects in the coating layer within the boundaries between the electrode metal and the BP flake, and also likely originates from the residue of PMMA during the deposition of electrodes that blocks the growth of hexylamine.

Such monolayer protection is effective not only for BP, and also for other layered 2D materials. Here to accelerate corrosion tests for *n*-hexylamine-coated 2D materials, we used harsh aqueous $H_2O_2$ or $KMnO_4$ solutions as etchants. In Table 1, we take the optical microscopy images during the corrosion exposure for each 2D material, including BP, $WS_2$, $WSe_2$, 1T'-$MoTe_2$, $WTe_2$, $TaS_2$, and $NbSe_2$. It should be noted that exfoliated BP, 1T'-$MoTe_2$, $WTe_2$, $NbSe_2$ and chemical vapor deposition (CVD)-grown single-layer $WS_2$ are known to be particularly susceptible to ambient corrosion and are readily attacked by solutions of $H_2O_2$. $WSe_2$ and $TaS_2$ are less vulnerable and require stronger oxidants for corrosion. *n*-hexylamine is proved to be effective in protecting these layered materials based on the comparison in optical image between uncoated and coated 2D materials after their exposure to the same etchants. A video of the corrosion retardation for BP is presented as Supplementary Movie.

Despite the fact that *n*-hexylamine is sturdy under various environments, it is still removable by certain organic acids. Presumably, the organic-media-supported protons can penetrate the hydrophobic alkyl layer, protonate the ionized surface $P{-}O^-$ groups, disrupting their electrostatic interaction with the alkylammonium cations. This removing protocol is effective both for the amine coating on BP and TMDs, without affecting the passivation oxidized layer and the materials underneath (Section 5 of SM).

## Discussion

Amines with low water solubility have long been known as efficient and reliable corrosion inhibitors for steels (*40*, *43*). It is found here that it also serves as an effective coating for 2D layered materials, by blocking water for the native thin oxide layer growing at the interface between the 2D material and the alkylamine coating. The photooxidation of bare BP starts with the synergetic effect of oxygen, water and light, where phosphorous transformed to a layer of acidic phosphorus species. The thin layer of acid then coarsens into a droplet, leaving a fresh phosphorous surface in contact with ambient air, and the oxidation process starts once again (*44*). *n*-hexylamine monolayer lowers the permeability of oxygen and strongly blocks the water molecules from directly contacting the oxide passivation layer and phosphorous. Although the first BP layer is still oxidized by $O_2$, it is isolated from ambient humidity by the hydrophobic alkyl monolayer, which prevents the water from dissolving this top native oxide that would have perpetuated the corrosion. Our experimental finding of the passivation effect on BP is consistent with the theoretical prediction that mere BP + $O_2$ reaction forming BP-$PO_x$ should be fully stable and self-limiting at ~1-2 nm if no moisture exists (*30*).

In summary, we have developed a strategy to effectively slow down the corrosion of BP by coating of alkylamine monolayer onto its surface. General applicability on a variety of other layered materials is also demonstrated. The alkylamine monolayer is robust in a range of chemical and thermal environments, including ambient air. The facile coating method can be implemented with many different substrates and is compatible with all linear alkylamines no shorter than *n*-butylamine, thus offering a platform for controlling the surface physics and chemistry of a rich tableau of 2D materials. Because of its simplicity, eco-friendliness and low cost, we envision it to be scalable and adaptable in various industrial configurations.

**Acknowledgements:** C. Su and Z. Yin would like to thank Philip Kim for granting lab access for the glove box enclosed AFM. C. Su would like to thank Greg Lin and Frank Zhao for helpful discussions. Z. Yin thanks Pablo Jarillo-Herrero for providing glove box. J.L., C.S., M.D., and L.S. acknowledge support by the Center for Excitonics, an Energy Frontier Research Center funded by the US Department of Energy, Office of Science, Basic Energy Sciences under award No. DE-SC0001088. J.L., C.S. and Z.Y. acknowledge support by NSF ECCS-1610806 and ANU Futures Scheme (Grant No. Q4601024). Q.B.Y. and G.S. acknowledge support in part by the MOST of China (Grant No. 2013CB933401), the NSFC (Grant No. 11474279), and the Strategic Priority Research Program of the Chinese Academy of Sciences (Grant No. XDB07010100). H.L. and J.H. acknowledge funding support provided by NSF under award No. 1453218. J.H.W. thanks the support from the Royal Society. J.K. and C.S. acknowledge support from the U.S. Army Research Office through the MIT Institute for Soldier Nanotechnologies, under Award No. 023674. T.Y., N.Z., and K.T. acknowledge support by Inter-University Cooperative Research Program of the Institute for Materials Research, Tohoku University(15G0031). **Data availability:** The raw/processed data required to reproduce these findings is available on request by personal communication.

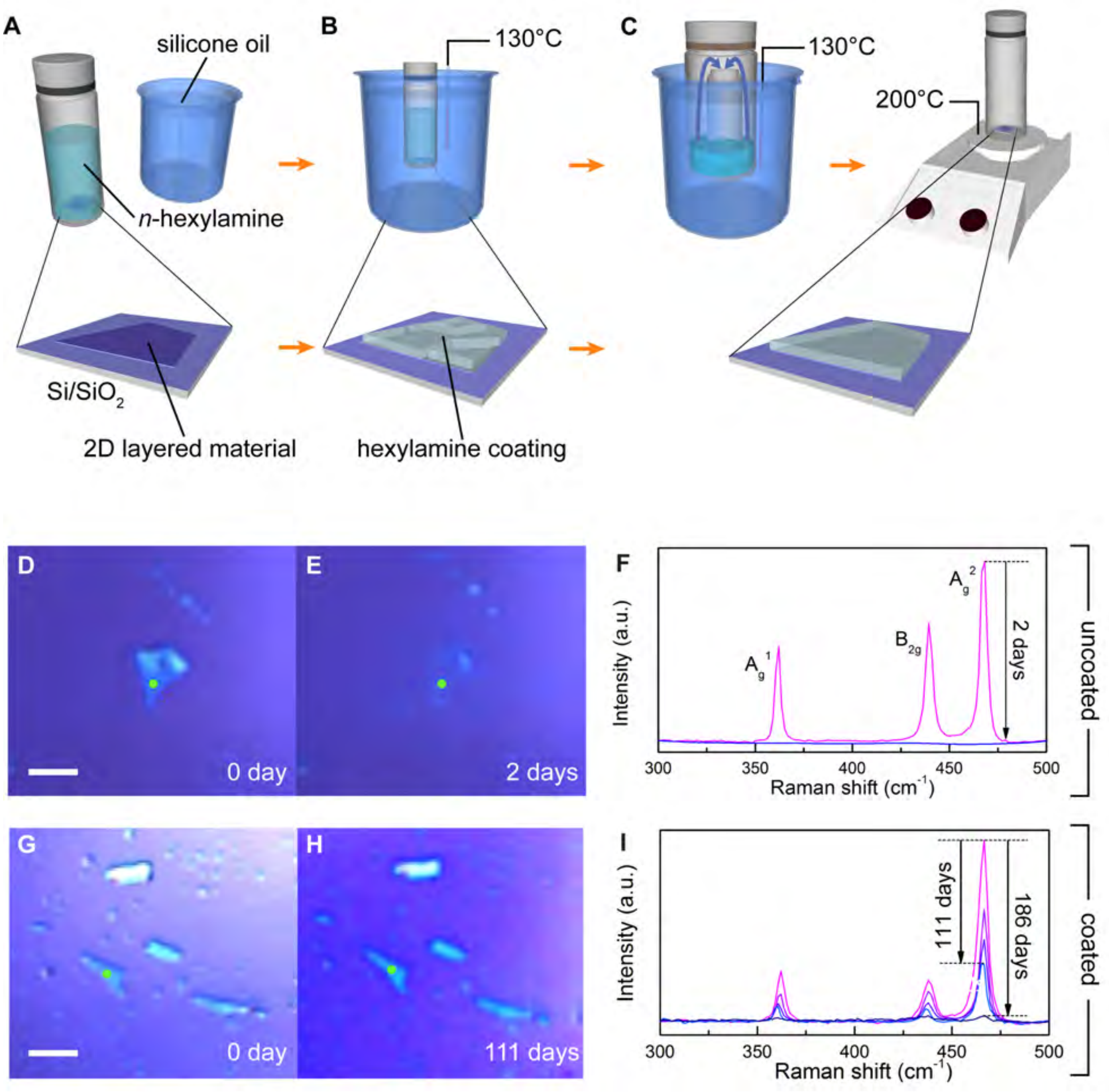

**Fig. 1. Coating *n*-hexylamine on BP flakes. A,** The layered materials on $SiO_2$/Si wafer are immersed in the liquid *n*-hexylamine contained in a glass vial, which is then capped and placed inside the silicone oil. **B,** In the first step, the oil bath is heated up to 130 °C and maintained for 20 minutes. Hexylamine is coated on the layered material with minor defects. **C,** In the second step, the sample is steamed in the amine vapor under 130 °C for 20 min, followed by an annealing in 200 °C for 30 min. This step is aimed for fixing the defects in *n*-hexylamine overlayer. **D, E,** Optical images of an exfoliated BP flake on $SiO_2$/Si wafer before aging (on 0

day), and after 2-day aging in ambient conditions where only the blurry marks of original flake could be identified. **F,** The corresponding Raman spectra at $\lambda$ = 532 nm of the sample shown in D and E. **G, H,** Optical images of *n*-hexylamine-coated BP flakes on $SiO_2$/Si before aging (on 0 day), and after 111-day aging under the same ambient conditions. **I,** The corresponding Raman spectra of the *n*-hexylamine-coated sample on the 0, 13, 41, 111, and 186 days. The green dots in D, E, G, and H are laser spot positions for repeated spectra acquisition. All samples are dried for 30 min at 120 $^{o}$C in air right after prepared. All Raman spectra shown above have been renormalized and calibrated to Si (reference) peak intensity. The scale bars are 5 μm.

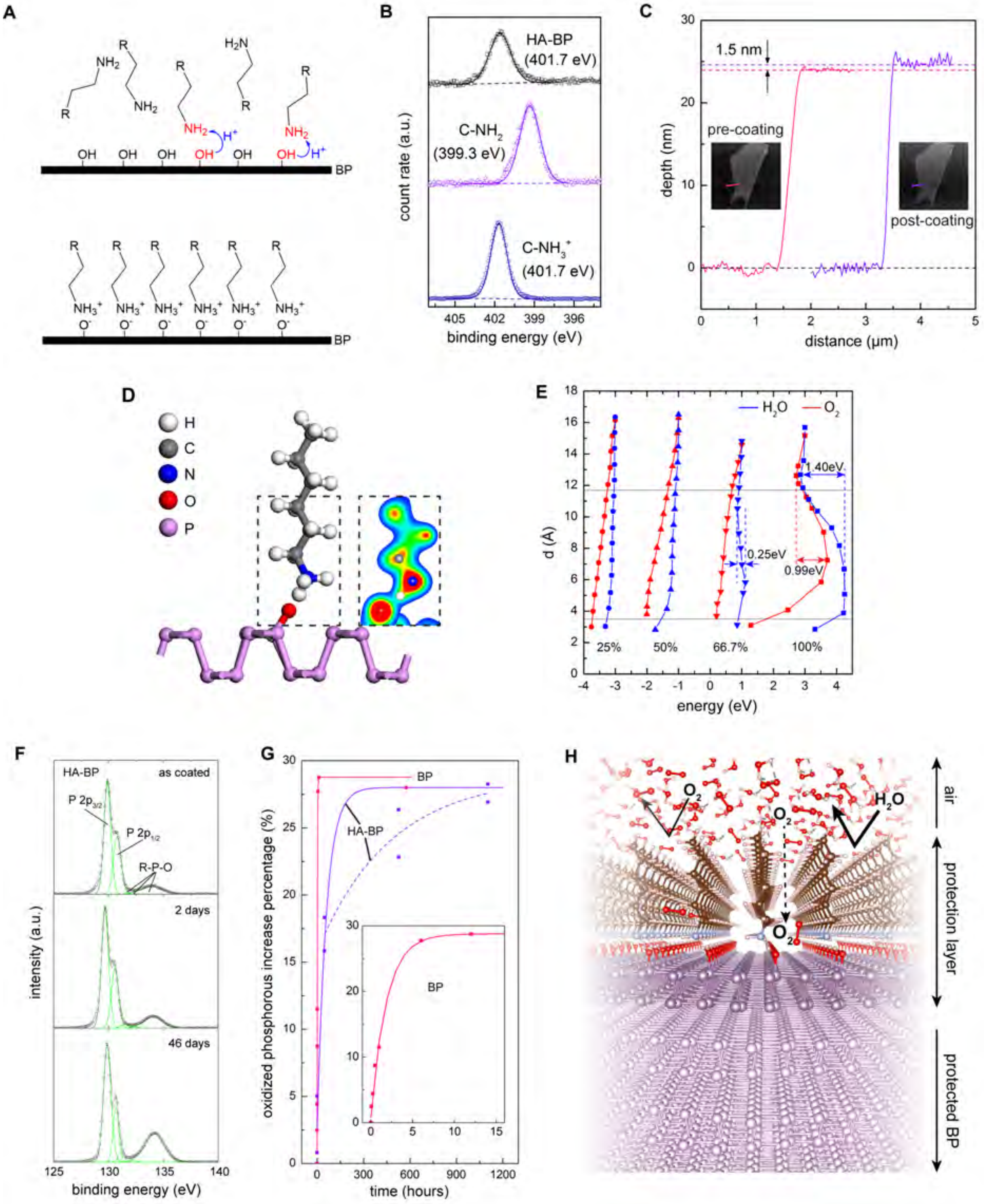

**Fig. 2. The mechanism of *n*-hexylamine coating on BP. A**, Proton transfer takes place during the coating process (upper figure) and the *n*-hexylamine monolayer is formed on BP after the coating process is done (lower figure). R- in the diagram refers to $C_4H_9$- when representing

hexylamine. **B**, XPS spectra of nitrogen 1s peaks on HA-BP, dodecylamine (C-$NH_2$), and $CH_3NH_3Cl$ (C-$NH_3^+$), proving that the amino group of *n*-hexylamine coated on BP is in ionic state –$NH_3^+$. **C**, The AFM data revealing the thickness of a BP flake with 24 nm before coating (pink line) and the thickness increment after hexylamine coating (violet line). **D**, The schematic structure of *n*-hexylamine adsorbed on BP, where red-, blue-, gray-, purple-, and white-colored balls represent oxygen, nitrogen, carbon, phosphorous, and hydrogen, respectively. Inset: the contour map of valence electron density on the plane containing O, N atoms and the H atom between them, which corresponds to the part marked by rectangle dashed line. **E**, The energy profile of $H_2O$ and $O_2$ molecules when penetrating through the hexylamine molecule layer. The y-axis is the distance between the bottom atom of $H_2O$ or $O_2$ and the surface of BP, denoted as *d*. Blue and red curves represent $H_2O$ and $O_2$ penetration processes, respectively. The four groups of curves represent different coverages of 25%, 50%, 66.7%, and 100% (detailed coverage definition illustrated in fig. S6 in SI), as marked. The horizontal grey lines are the locations of the top and the bottom of hexylamine molecules. **F,** The P 2p peaks and oxidized phosphorous species (R-P-O) of XPS curves on HA-BP measured as coated, after 2 days, and after 46 days. **G,** The phosphorous oxide concentration as a function of time between *n*-hexylamine-coated (violet triangles) and un-coated BP samples (pink squares). Inset: a blow-up of the un-coated sample data between 0-15 hours. Both data sets are fitted with exponential curves. The pink and violet solid lines are fittings of the scattered data pointing to the un-coated and HA-coated samples, respectively. Note that the oxidation of HA-BP is significantly slow down starting from 100 hours, so a second curve fitting is marked (dashed violet line). **H**, Schematic illustration of the structure of BP after coated by *n*-hexylamine. The first layer of BP is oxydized and forms a part of protective layer together with the *n*-hexylamine coating. The surface protective layer (hexylammonium + first layer oxidized BP) protects the rest of BP underneath.

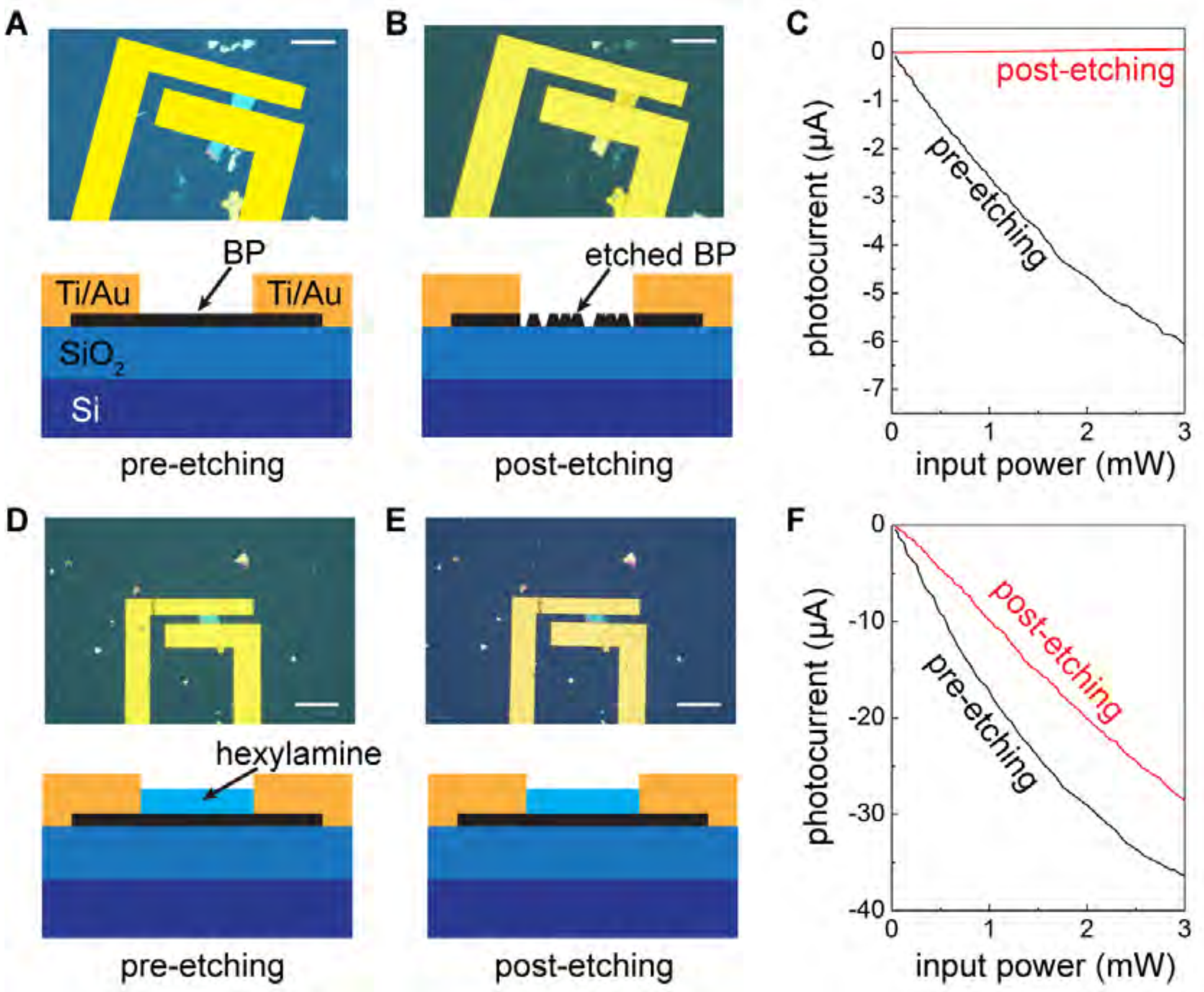

**Fig. 3. Photodetectors and etching test.** **A, B,** The uncoated BP devices before and after etching by an aqueous solution of 30 wt% $H_2O_2$. The optical images of the devices are shown in the upper row and the corresponding schematic layouts are shown in the lower row. The Ti/Au electrode in (a) is marked by yellow stripes after the picture is taken. **C,** Photocurrent as a function of input optical power under zero bias voltage. **D-F,** *n*-hexylamine-coated-device counterparts of sub-figures (a-c). The scale bars are 10 μm.

**Table 1. Protection of various 2D materials with *n*-hexylamine coatings.** BP, $WS_2$, 1T'-$MoTe_2$, $WTe_2$, $WSe_2$, $TaS_2$, and $NbSe_2$ were coated with *n*-hexylamine and dipped inside etchants of $H_2O_2$ or $KMnO_4$ solution (depending on the respective material reactivity) as an accelerated lifetime test. The uncoated counterparts were processed in parallel with the coated parts under identical etching conditions. Scale bars represent 10 µm.

| **material** / etching with etching time | **bare** | | **coated** | |
|---|---|---|---|---|
| | **before exposure** | **after exposure** | **before exposure** | **after exposure** |
| **BP (exfoliated)** / 20sec in $H_2O_2$ (30 wt. % in $H_2O$) | 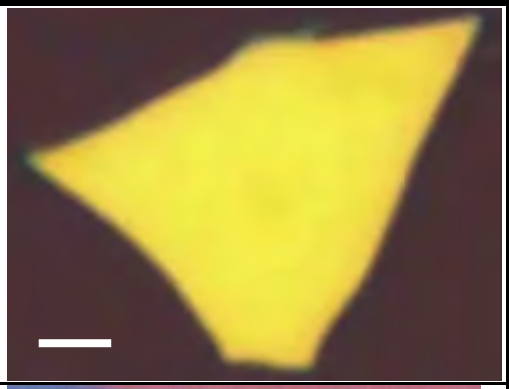 | 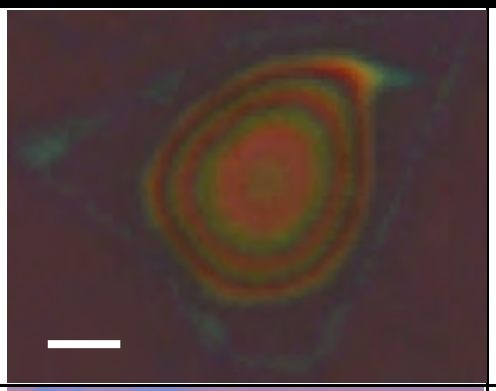 | 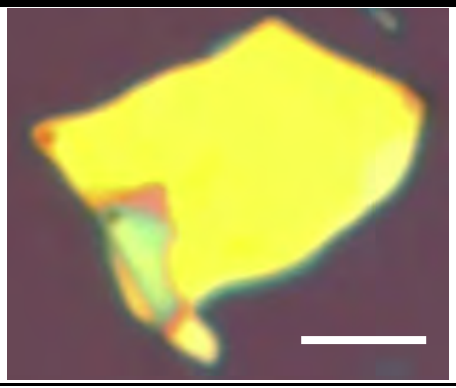 | 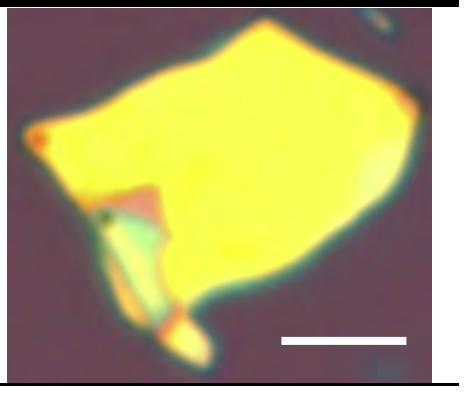 |
| **$WS_2$ (CVD, monolayer)** / 5sec in $H_2O_2$ (30 wt. % in $H_2O$) | 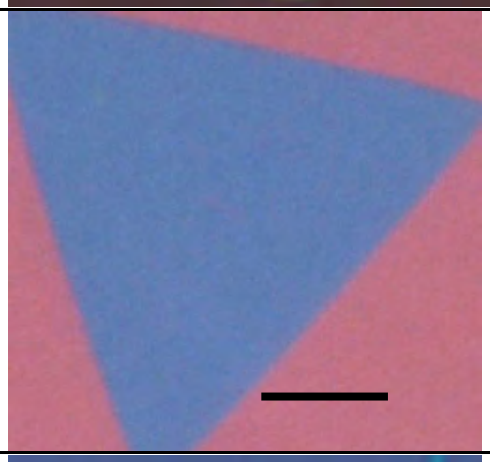 | 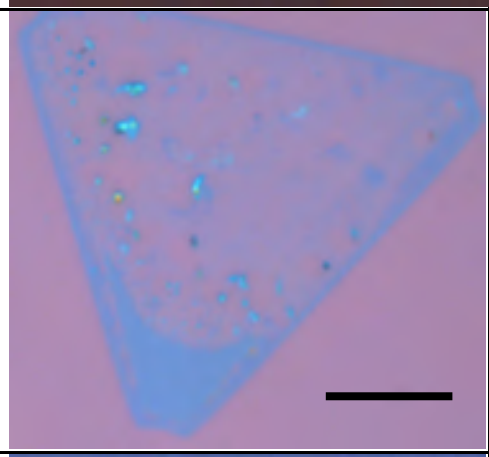 | 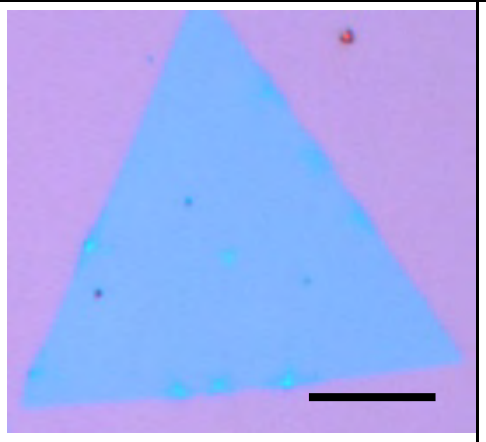 | 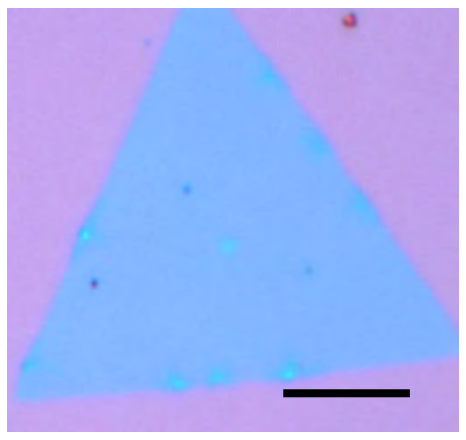 |
| **1T'-$MoTe_2$ (exfoliated)** / 10sec in $H_2O_2$ (30 wt. % in $H_2O$) | 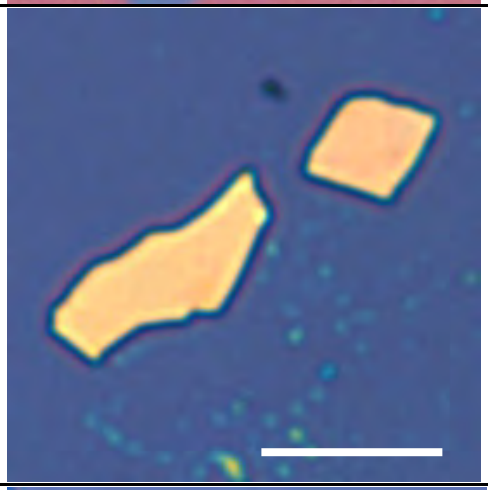 | 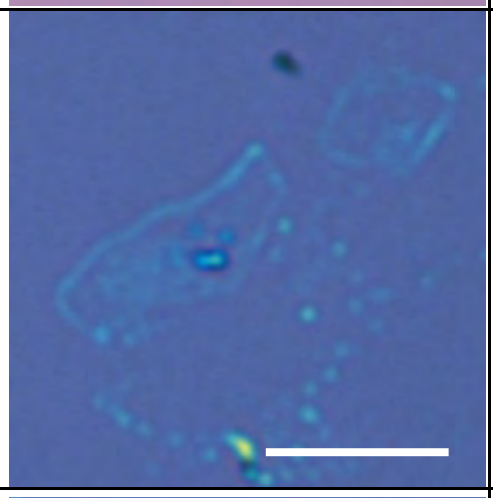 | 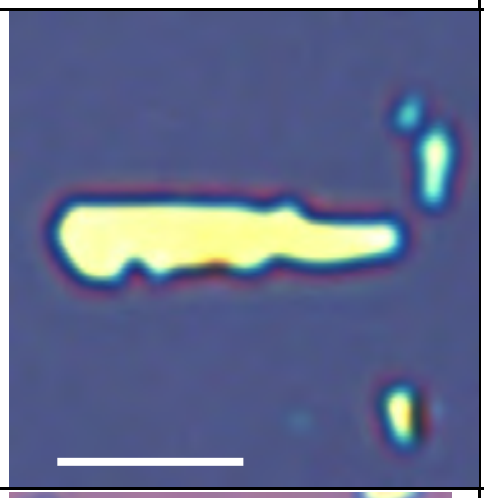 | 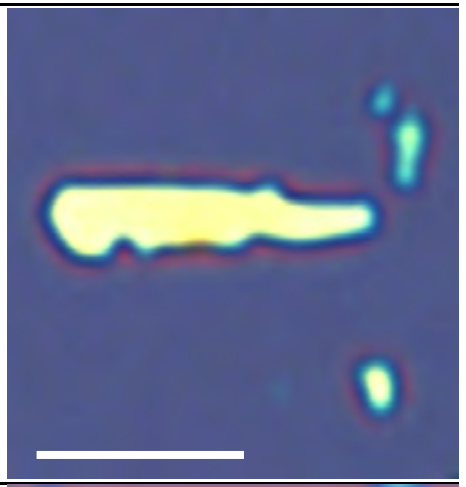 |
| **$WTe_2$ (exfoliated)** / 30sec in $H_2O_2$ (30 wt. % in $H_2O$) | 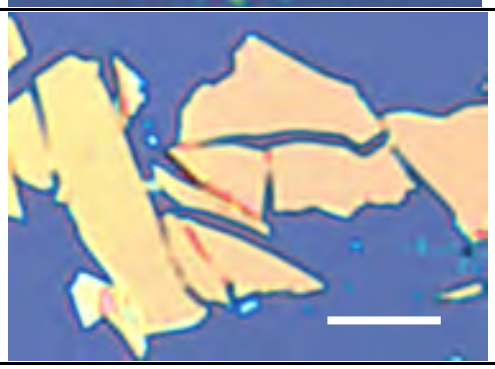 | 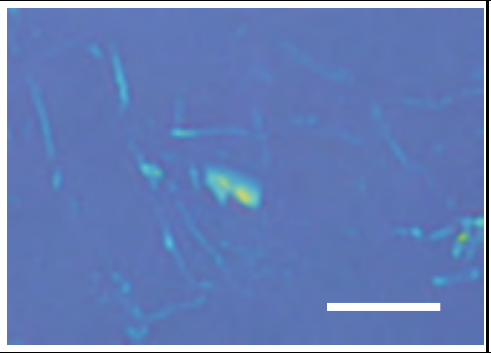 | 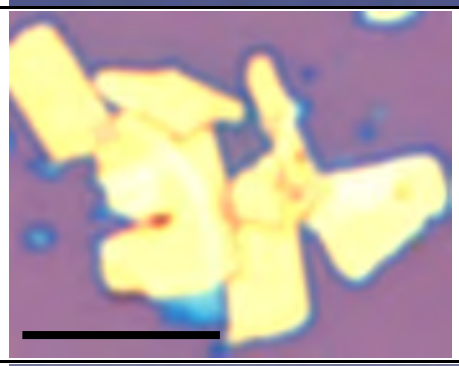 | 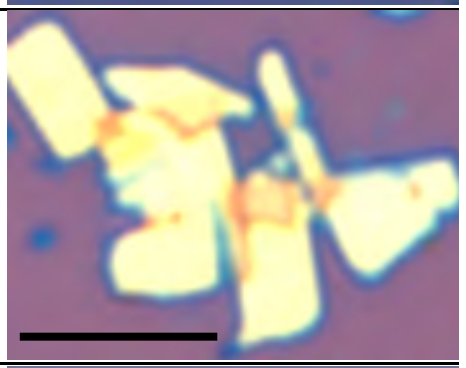 |
| **$WSe_2$ (exfoliated)** / 1min in $KMnO_4$ (0.02mol/L in $H_2O$) | 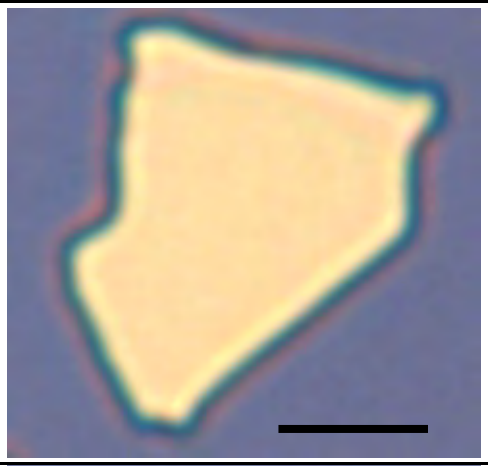 | 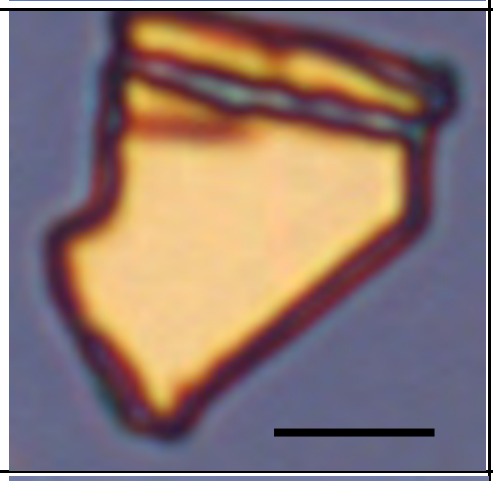 | 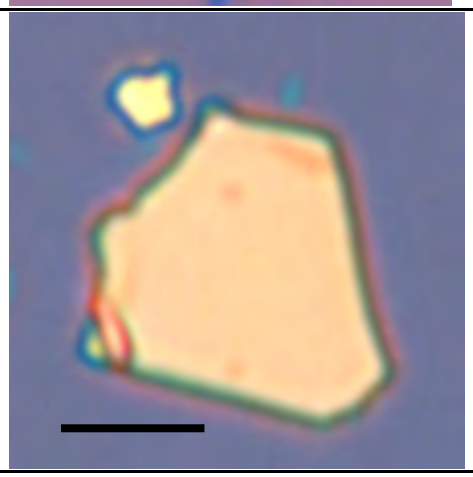 | 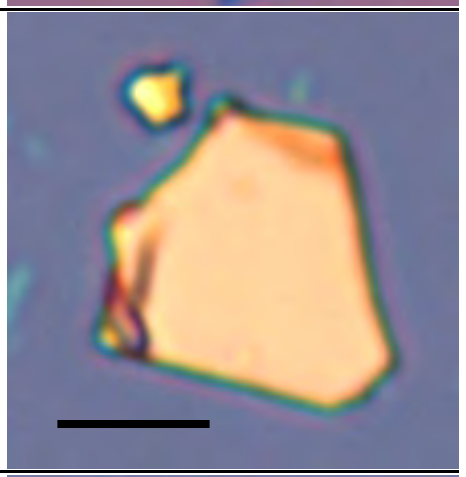 |
| **$TaS_2$ (exfoliated)** / 1min in $KMnO_4$ (0.01mol/L in $H_2O$) | 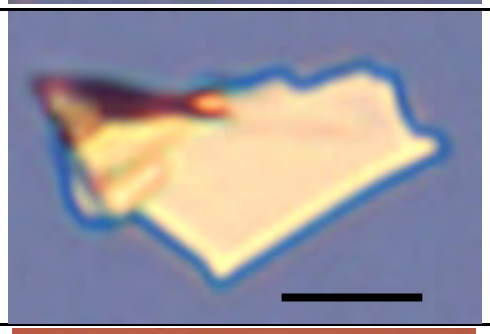 | 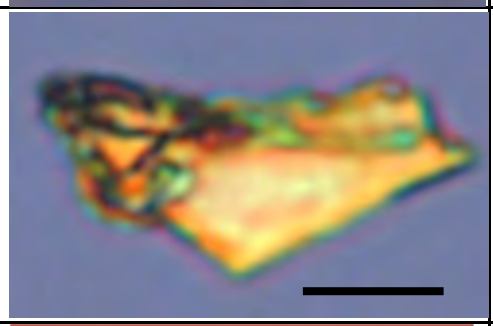 | 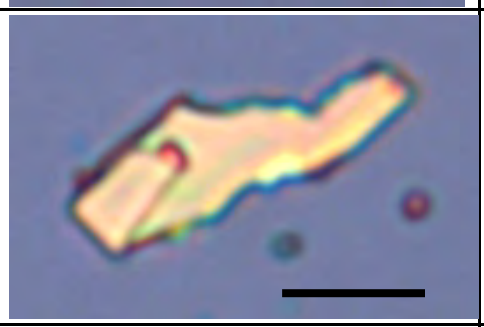 | 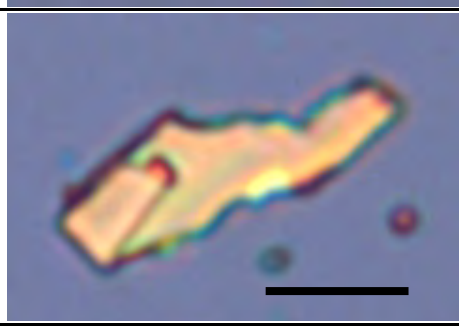 |
| **$NbSe_2$ (exfoliated)** / 20sec in $H_2O_2$ (30 wt. % in $H_2O$) | 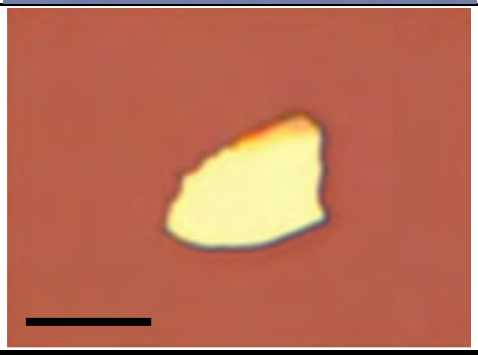 | 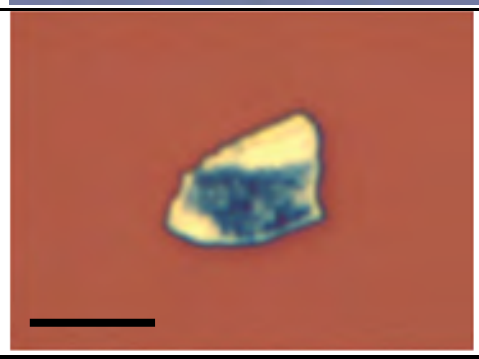 | 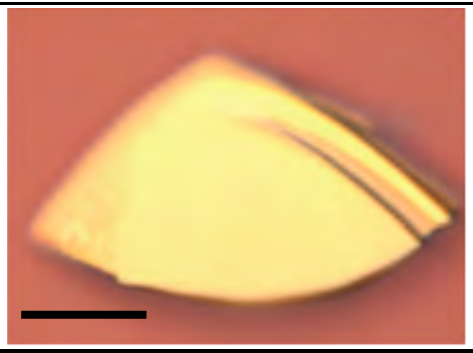 | 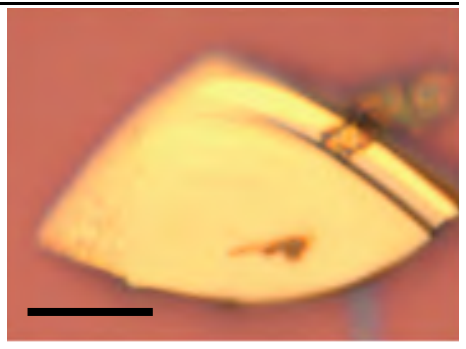 |

Supplementary Materials

# Molecular monolayer stabilizer for multilayer 2D materials

## Table of contents

## 1. Coating methods and protection testing

I. Coating of n-hexylamine ($C_6H_{13}NH_2$) onto unstable 2D flakes of black phosphorous (BP), $MoTe_2$, $WTe_2$, $WSe_2$, $TaS_2$, and $NbSe_2$

In our experiment, *n*-alkylamine (Sigma-Aldrich, 99%) was used as purchased. The two dimensional (2D) crystals are mechanically exfoliated and transferred or directly chemical vapor deposition (CVD) onto a piece of silicon wafer (with a 190 nm $SiO_2$ surface layer), denoted as 2D/$SiO_2$/Si, after $SiO_2$/Si substrates were washed in water, isopropyl alcohol (IPA) and acetone, respectively, by sonication for 10 mins, followed by the annealing in air for 30 min at 200 °C to remove the absorbed water on surface. The exfoliation was done with Scotch Tape in a glove box for BP, but in the air for other four 2D materials: $MoTe_2$, $WTe_2$, $WSe_2$, $TaS_2$, and $NbSe_2$. In the following, coating of *n*-hexylamine onto BP is taken as the example to introduce the whole coating procedure.

The whole coating process can be divided into two steps, which is performed in the Acrylic glove box, as schematically explained in the Fig. 1. Such glove box can maintain a certain level (~30 ppm) $O_2$ and $H_2O$, which is necessary for uniform oxidation and hydroxylation of BP surface layer during amine growth. In brief, 2D/$SiO_2$/Si samples were completely immersed in excess amount (2 - 10 mL, depending on the size of reactor vial or petri dish) of *n*-hexylamine contained in a glass vial or petri dish, covered with a cap. Such vial or petri-dish based reactor was immersed into a silicone oil bath or sitting on a hotplate which will be heated up for the first step growth. This step growth was maintained about 20 min at 130 °C and then cooled down to room temperature (RT) and kept for half an hour to complete the first step growth. After this, samples were taken out and rinsed with hexane to remove the attached amine residues. For the second step growth, the samples under heating at 130°C will be steamed in the amine vapor for about 20 min, with the subsequent cooling down to RT for another half an hour. After gently rinsed with hexane and dried, the samples were transferred to another glove box with low $O_2$ and $H_2O$ levels (<0.1ppm) to be sealed in a glass vial for the final simple post-growth annealing at 200 °C for 30 min. After cooling down, the sample was then ready for characterization and testing.

*n*-hexylamine coating parameters were optimized step by step based on BP (Table S1). The similar parameters are applied for *n*-hexylamine coating onto other 2D materials (Table 1 in main text), and the parameters are optimized for other amine molecules with different carbon chain lengths onto BP (Table S2 and S3).

II. Coating of *n*-hexylamine onto relatively stable $WS_2$

The coating protocols are slightly different from above. As-purchased *n*-hexylamine was directly used for the coating without further purification. $WS_2$ samples were grown onto $SiO_2$ (300 nm)/Si substrates by CVD method.[1] Subsequent deposition of *n*-hexylamine was realized by one-step growth. In order to drive oxygen out of *n*-hexylamine before the growth, *n*-hexylamine was boiled at ~130 °C for 30 min in air. Then, the $WS_2/SiO_2/Si$ samples were immediately immersed into boiling *n*-hexylamine carefully and covered with the cap (note: the cap should not be fully tightened to avoid high pressure building-up in the bottle that can lead to explosion). After growth and subsequent cooling down to room temperature, the samples were collected and gently rinsed with hexane followed by immediate drying with $N_2$ gas blowing. Similarly, as the two-step growth for other 2D material, a simple post-growth annealing was also applied for amine coated $WS_2/SiO_2/Si$ at 200 °C for 30 min in the argon and then the samples are ready for testing and characterization.

III. Coating other alkylamines and control molecules onto BP

Besides *n*-hexylamine (*n*-$C_6H_{13}NH_2$), BP crystals were also coated with other linear alkylamines with different carbon chain lengths, including *n*-butylamine (*n*-$C_4H_9NH_2$), *n*-pentylamine (*n*-$C_5H_{11}NH_2$), *n*-octylamine (*n*-$C_8H_{17}NH_2$), *n*-decylamine (*n*-$C_{10}H_{21}NH_2$), *n*-undecylamine (*n*-$C_{11}H_{23}NH_2$). The main difference in coating process for these amine molecules lies in the coating temperatures, which are normally set to be below or close to the boiling point of each amine molecule for safety considerations. The coating parameters are summarized in the table S2 below. After the coating, all these amine-coated BP samples were tested using $H_2O_2$ as etchant, where the protection effect is evaluated by comparing the optical microscope images before and after the $H_2O_2$ etching (table S3).

Furthermore, two control experiments were performed in terms of the molecule types used for coating. First, we tested benzylamine ($C_6H_5CH_2NH_2$), which is a non-linear alkylamine. Non-linear benzylamine is expected to be challenging to form a high-coverage dense monolayer on the surface of 2D material as there exists un-coverable gap between the benzene ring structures, thus affecting the protection as demonstrated below. Second, amino-group-free alkane molecule, i.e. *n*-hexane (*n*-$C_6H_{14}$), was tested on BP. As presented in table S3, *n*-hexane coating does not have any protection capability. This further consolidates our proposed monolayer link model between alkylamine and BP as discussed detailed in the main text.

IV. Protection testing results for alkylamines coated BP

**Table S1**. **Optimization of coating parameters of *n*-hexylamine on BP.** The etching method with $H_2O_2$ (30% wt. in $H_2O$) etchant/oxidant is as follows: dip BP into $H_2O_2$ for 20 sec, remove BP from $H_2O_2$, and leave dipped BP for 2 mins in air. Scale bars are 20 μm.

| | **Before applying oxidant** | **After applying oxidant** |
|---|---|---|
| **Pure fresh BP** <br> (not protected) | 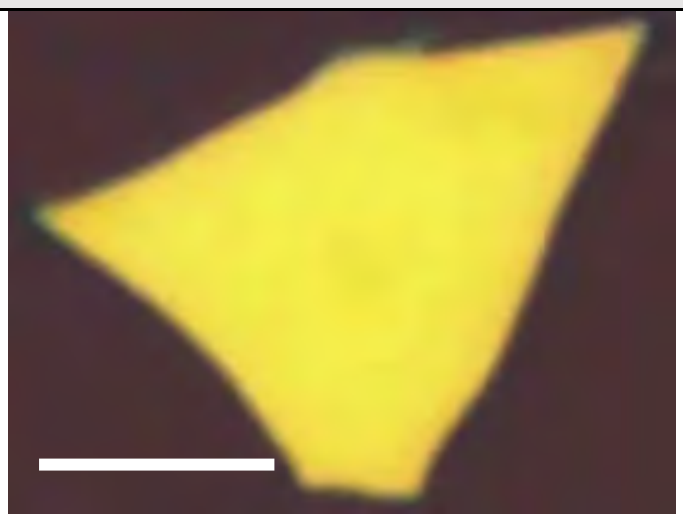 | 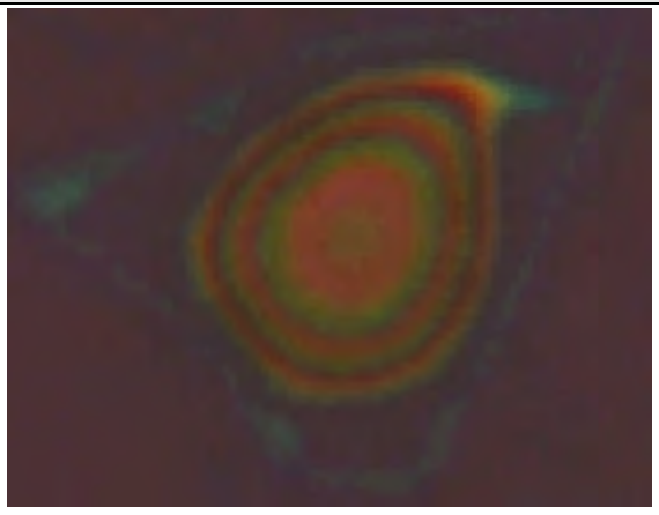 |
| ***n*-hexylamine-BP**: Immersed in *n*-hexylamine for 6 days at RT <br> (not protected) | 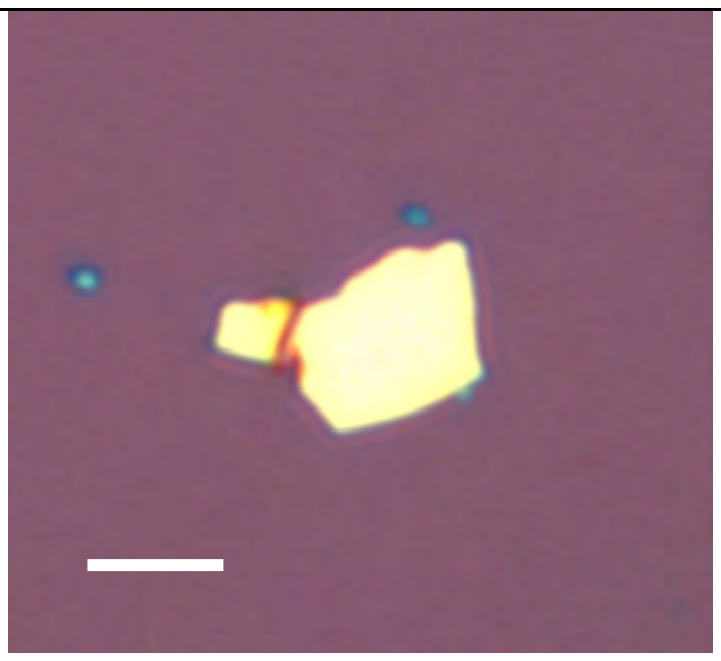 | 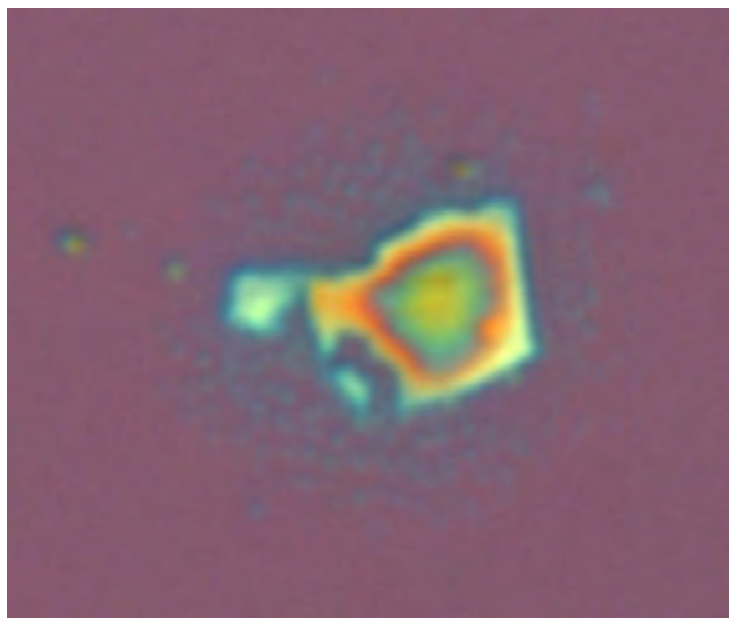 |
| ***n*-hexylamine-BP:** 70 °C & 20 mins <br> (not protected) | 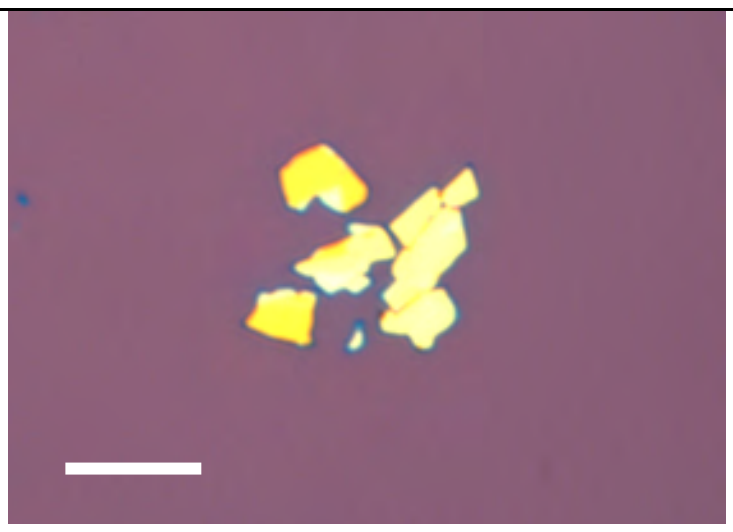 | 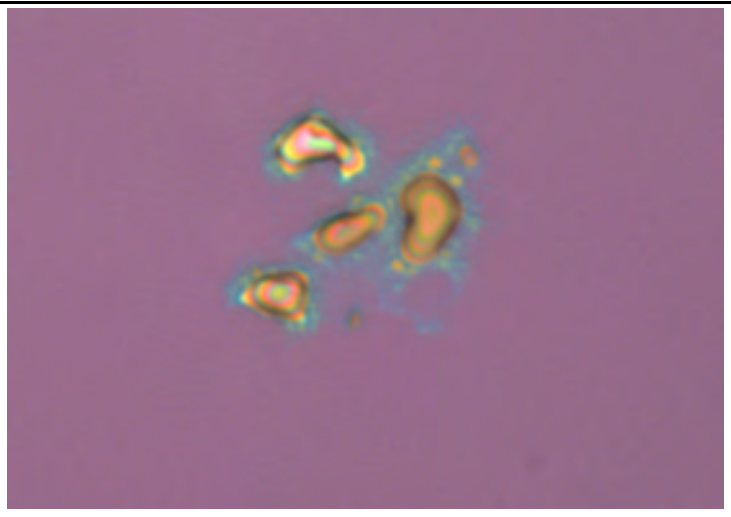 |
| ***n*-hexylamine-BP:**<br>90 $^{o}$C & 20 mins<br><u>(not protected)</u> | 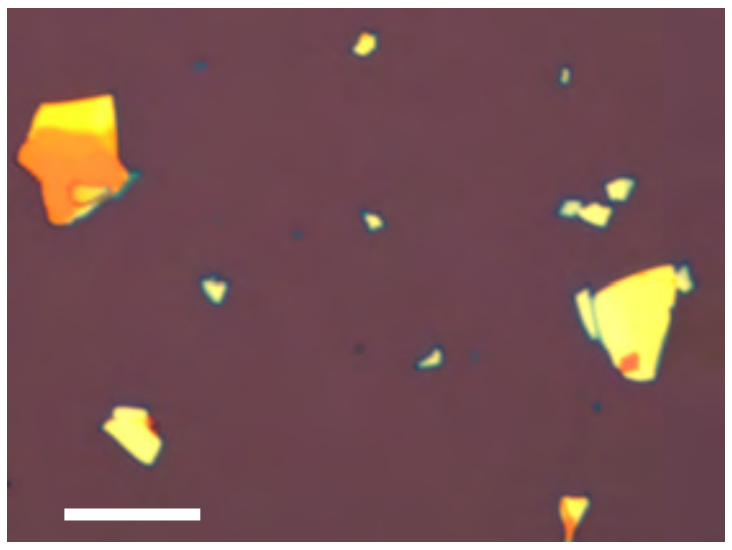 | 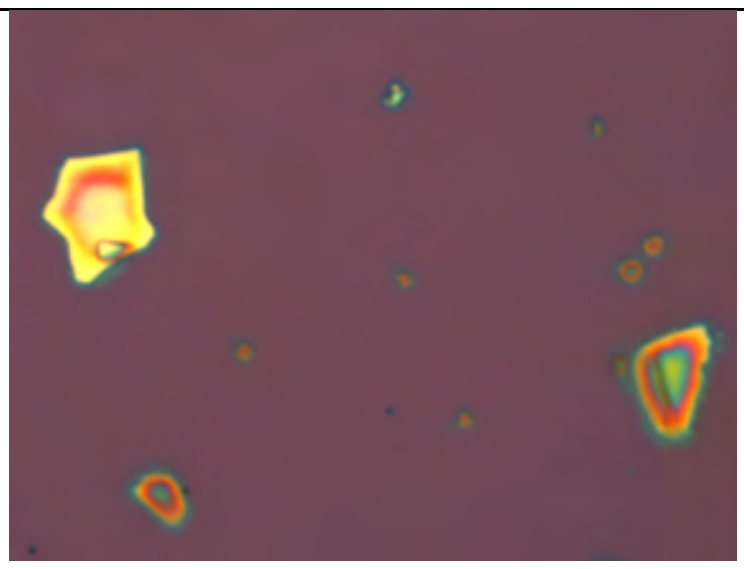 |
| ***n*-hexylamine-BP:**<br>110 $^{o}$C & 20 mins<br><u>(not protected)</u> | 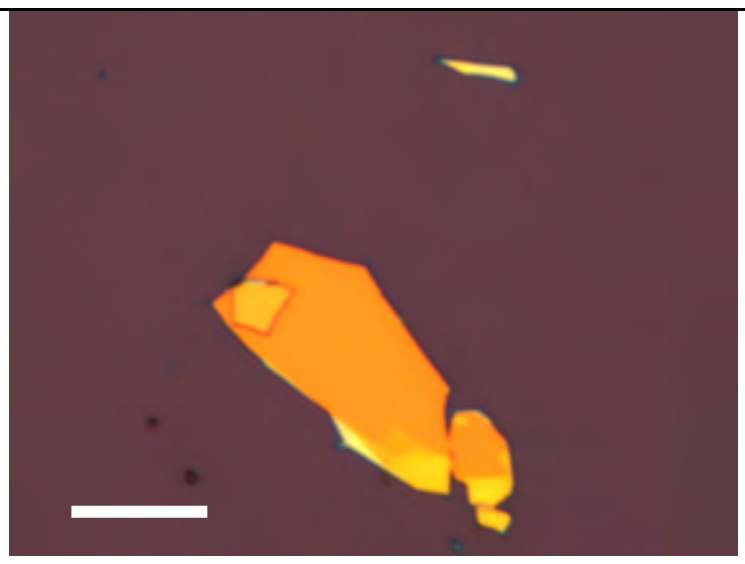 | 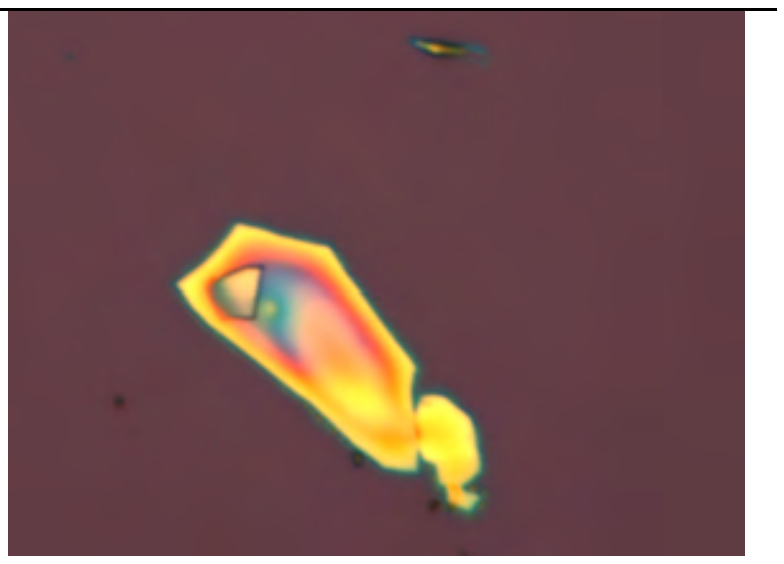 |
| ***n*-hexylamine-BP:**<br>130 $^{o}$C & 20 mins<br><u>(partially protected)</u> | 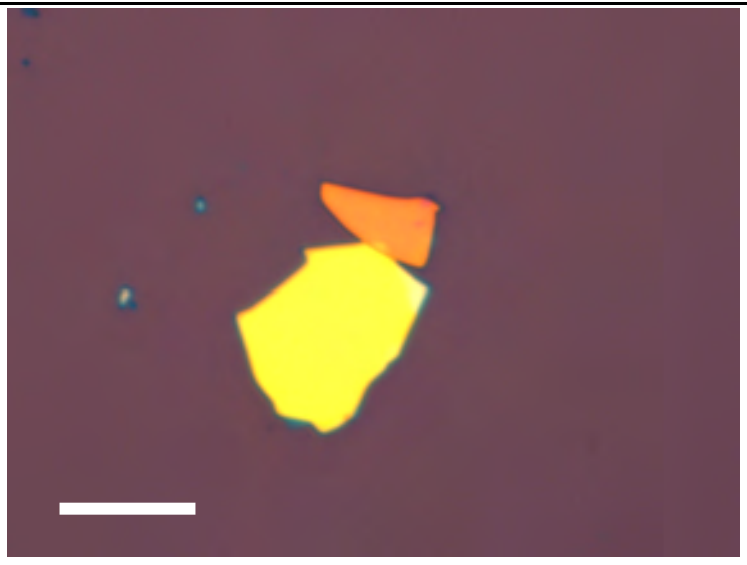 | 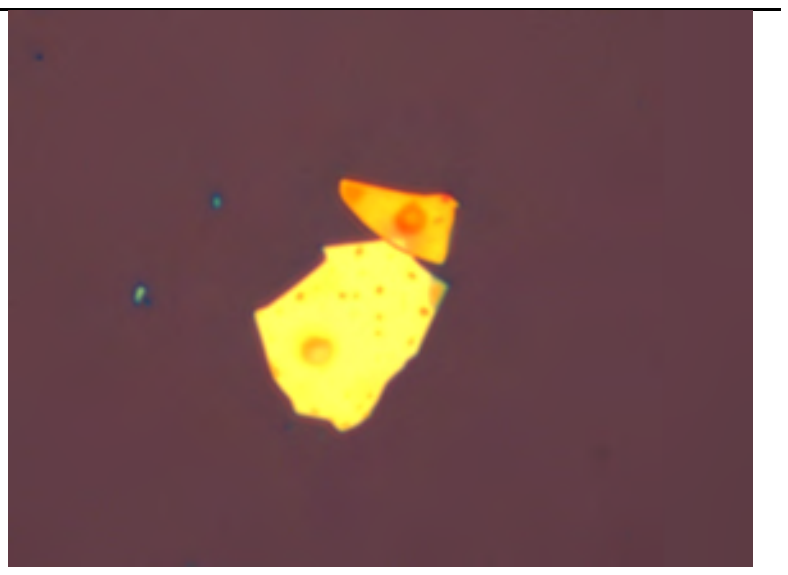 |
| ***n*-hexylamine-BP:**<br>two-step coating<br>aforementioned<br><u>(protected)</u> | 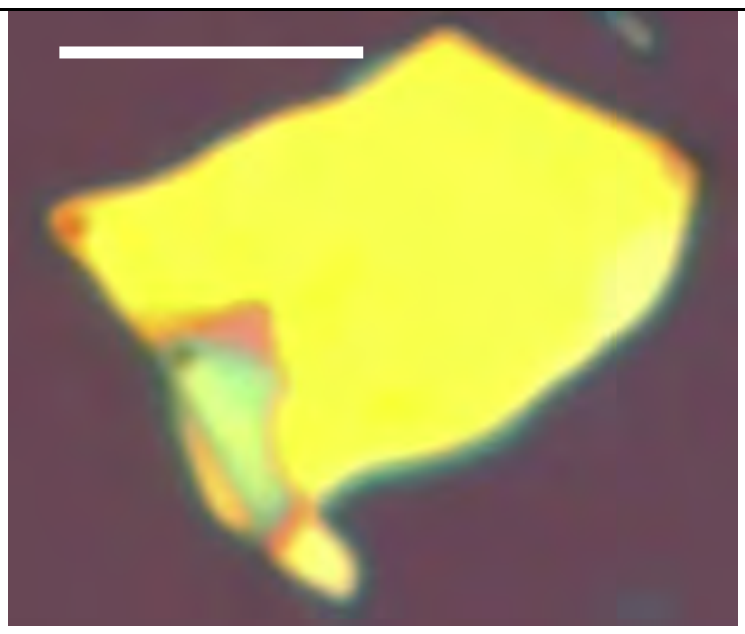 | 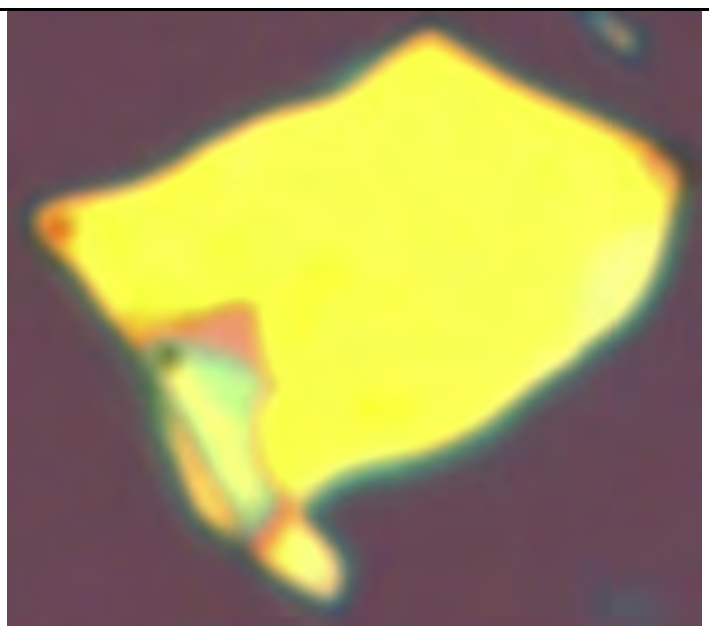 |

**Table S2**. **Protection testing for different alkylamines and hexane.** The optical microscope images were taken before and after etching/oxidation of BP flakes with the same etching method as described above.

| | **Coating temperature (°C)** | **Before applying oxidant** | **After applying oxidant** |
|---|---|---|---|
| **Pure fresh BP** (not protected, the same as Table S1) | **No treatment** | 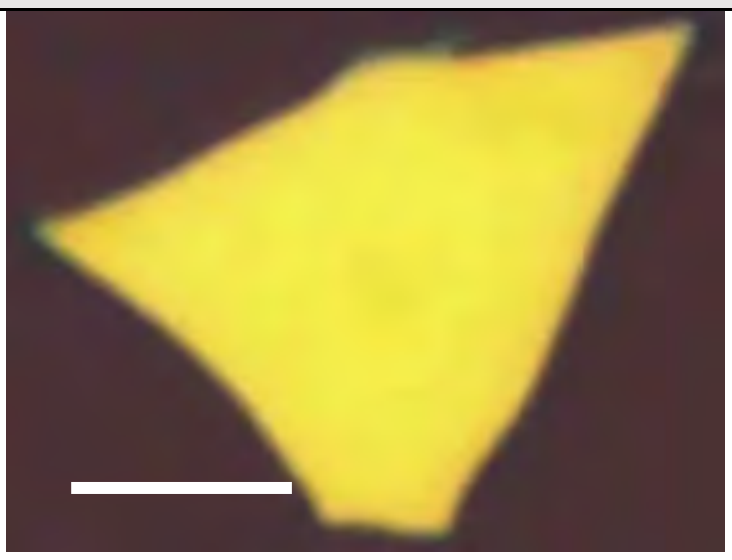 | 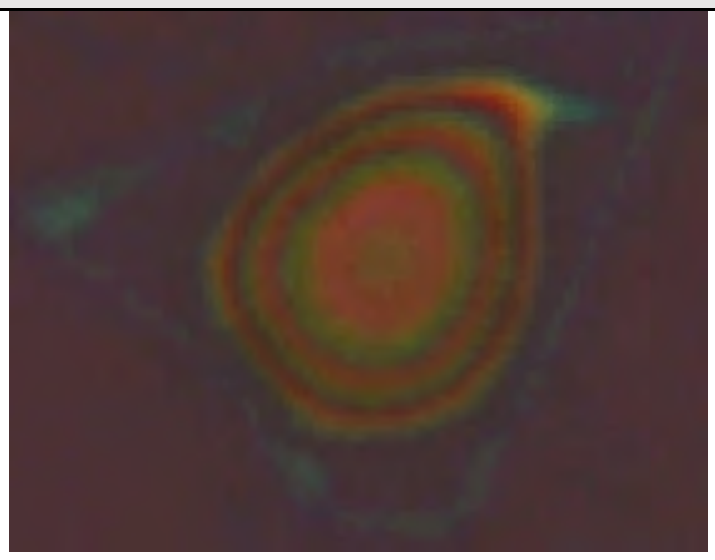 |
| ***n*-butylamine-BP** | **90** | 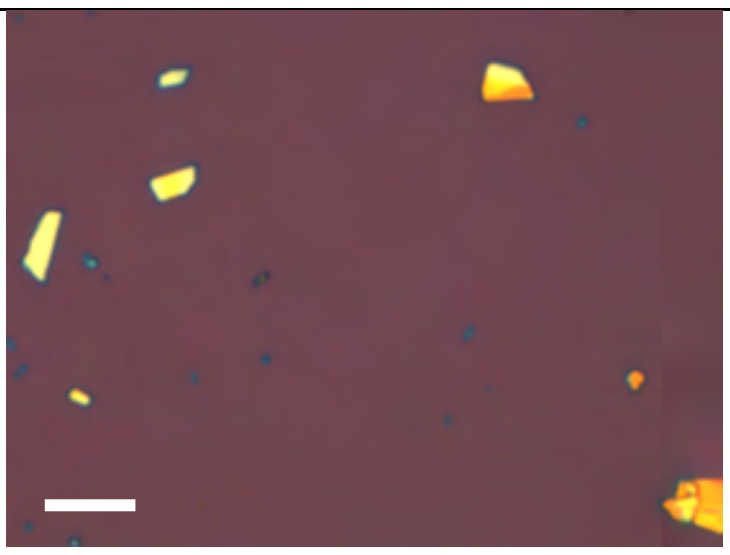 | 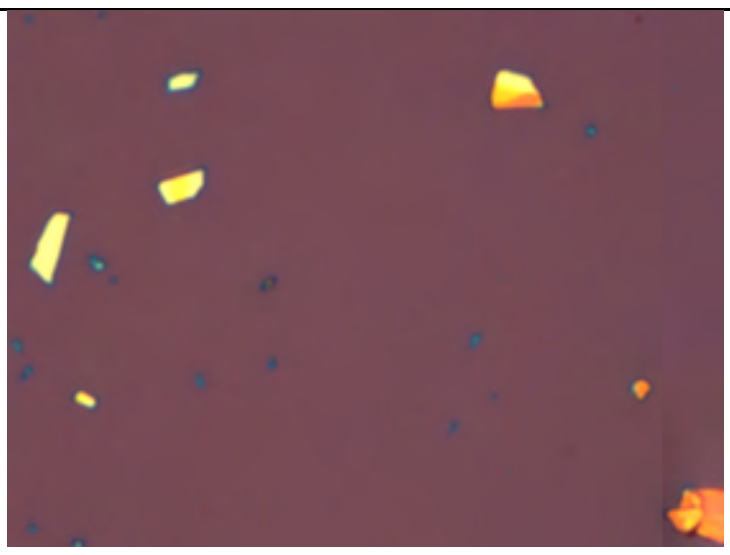 |
| ***n*-pentylamine-BP** | **110** | 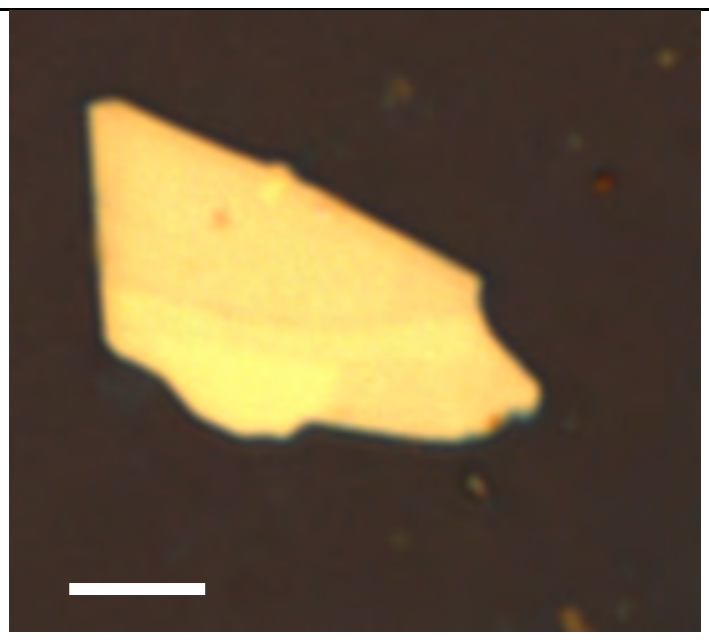 | 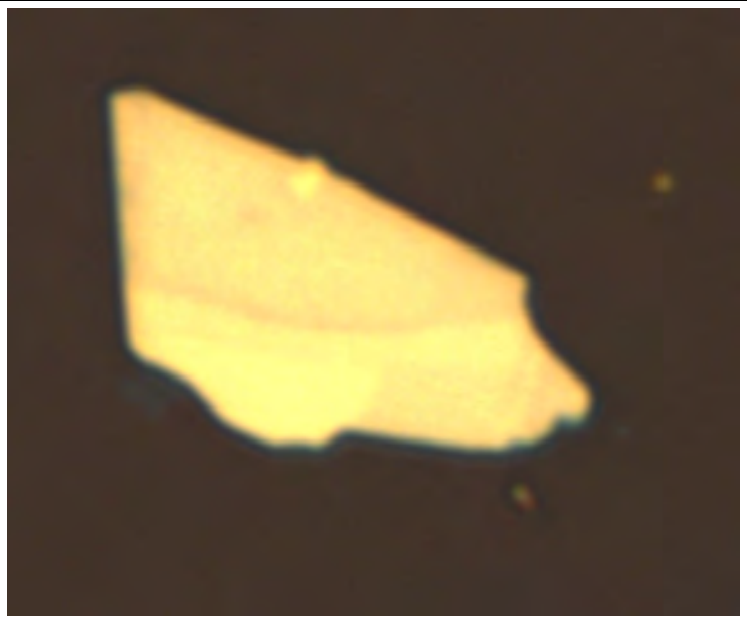 |
| ***n*-hexylamine-BP** | **130** | 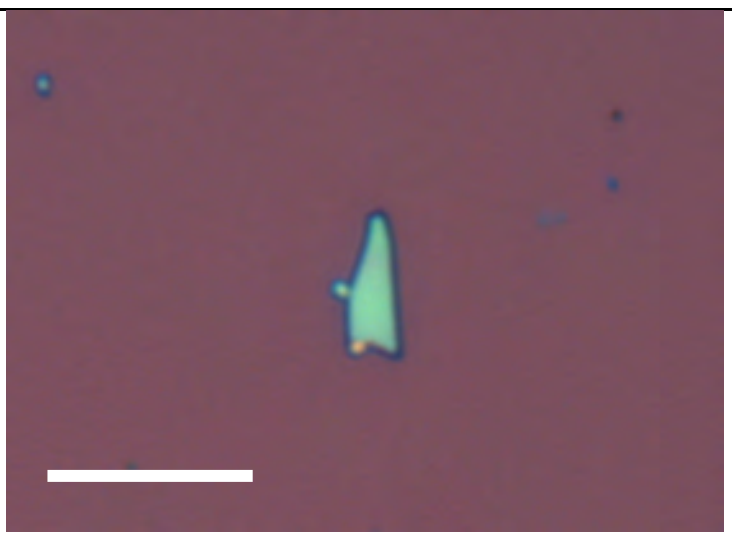 | 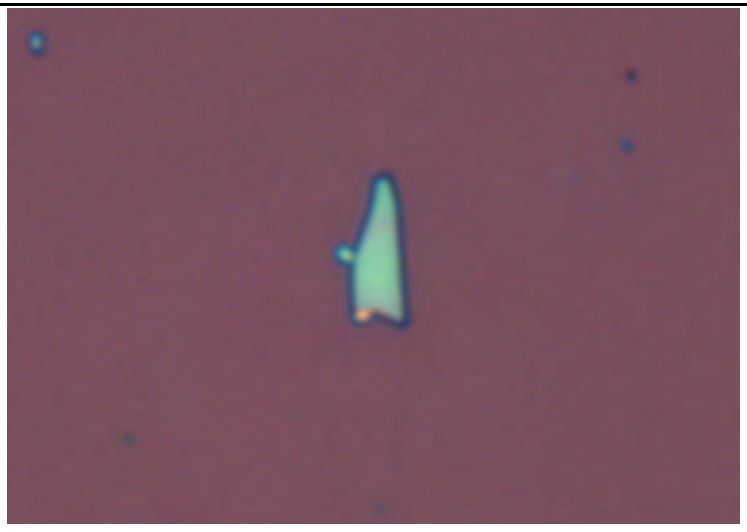 |
| ***n*-octylamine-BP** | **140** | 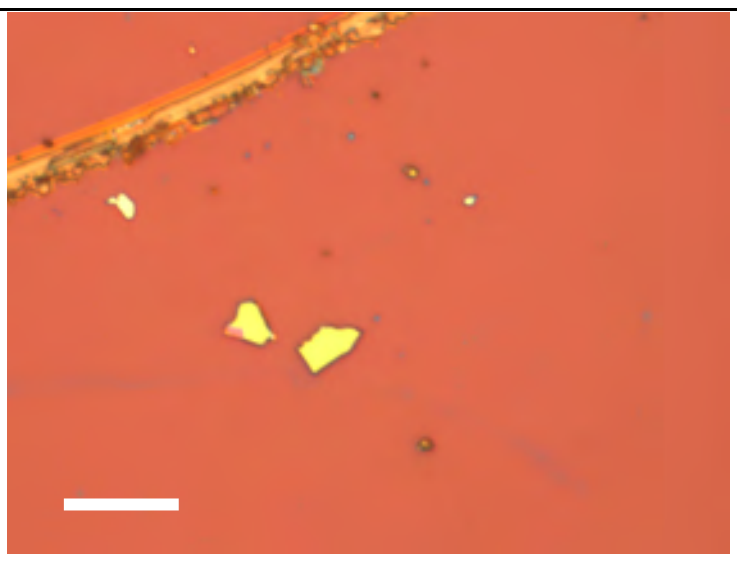 | 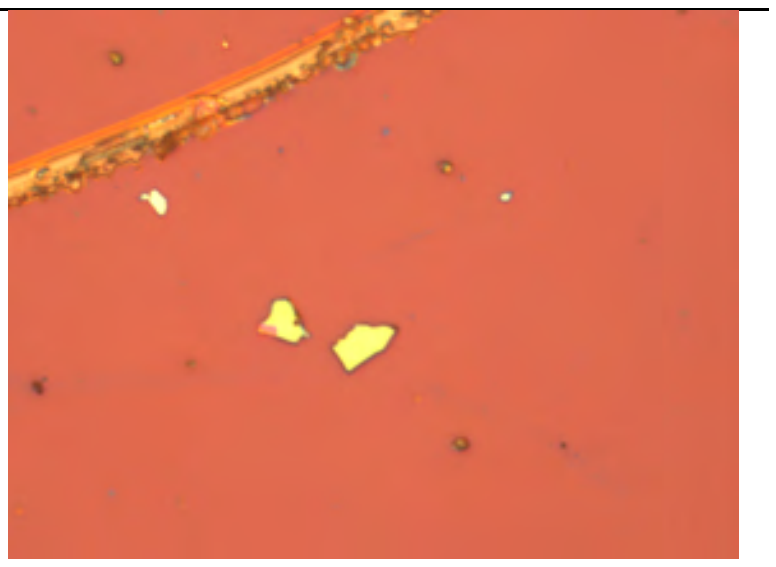 |
| | **160** | 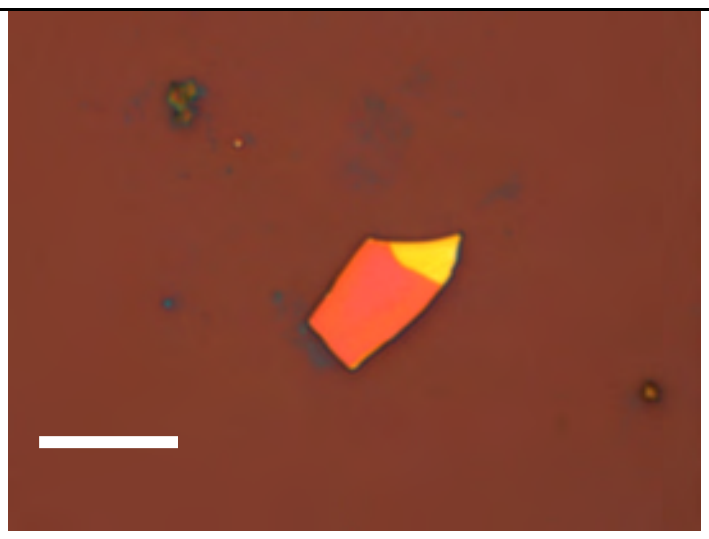 | 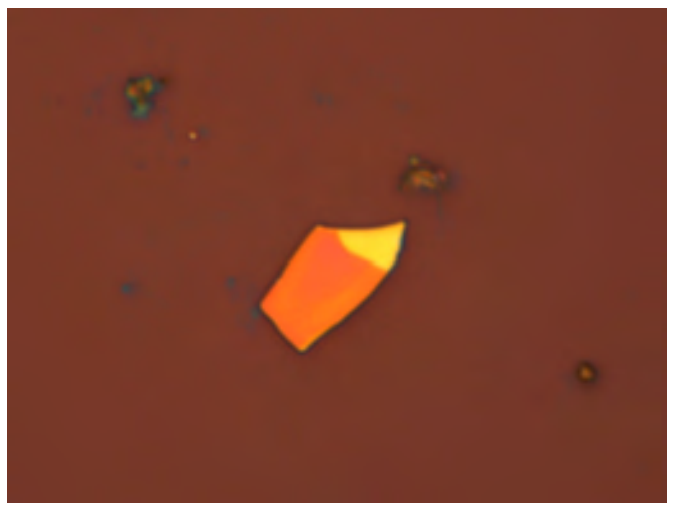 |
| ***n*-decylamine-BP** | **120** (partially protected) | 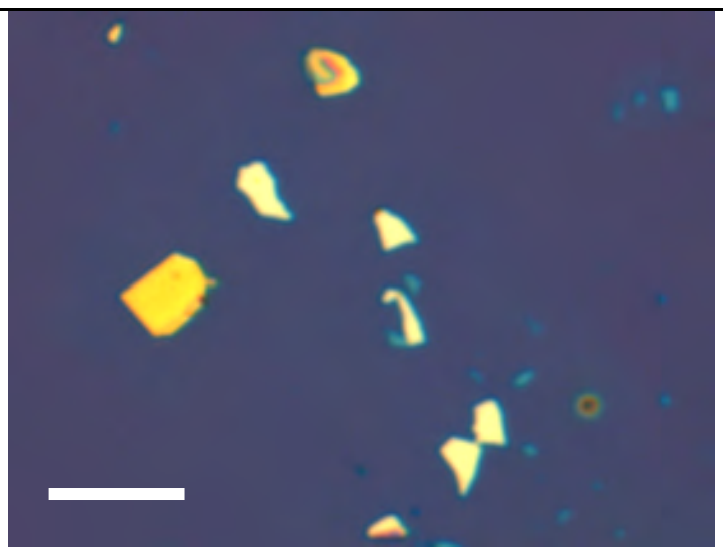 | 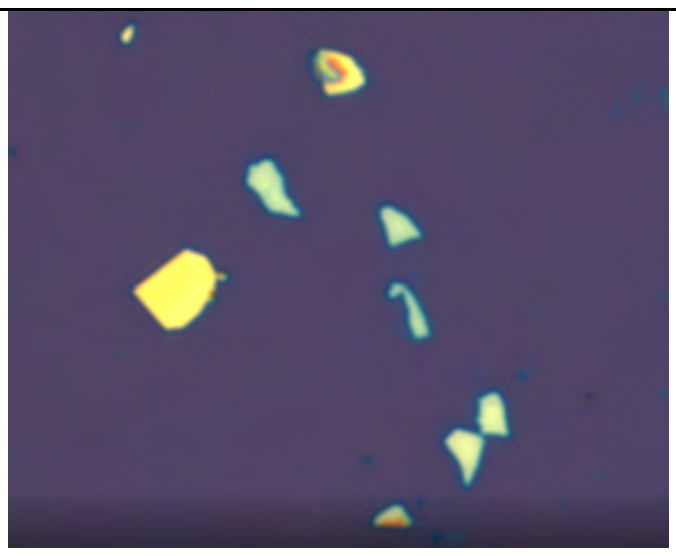 |
| | **150** | 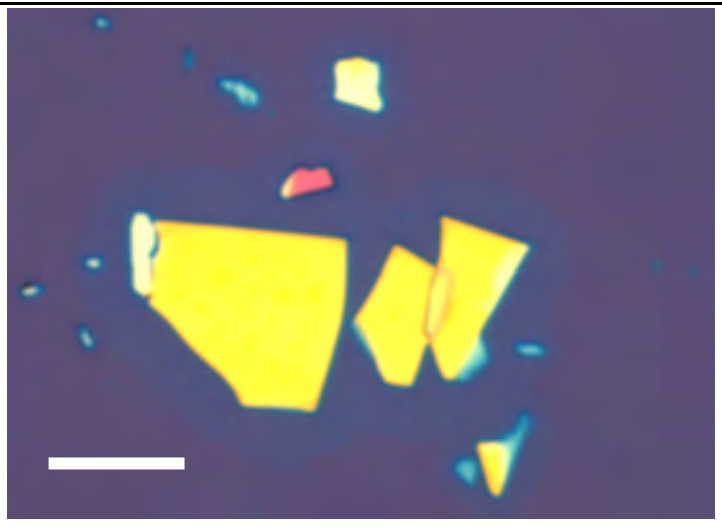 | 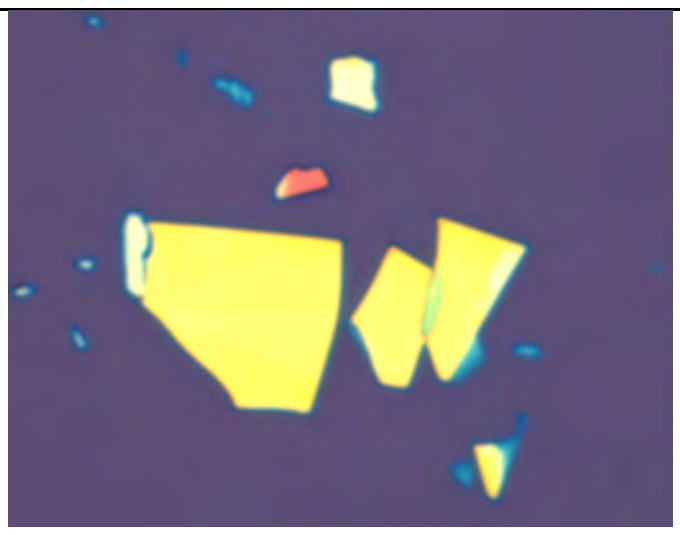 |
| | **180** | 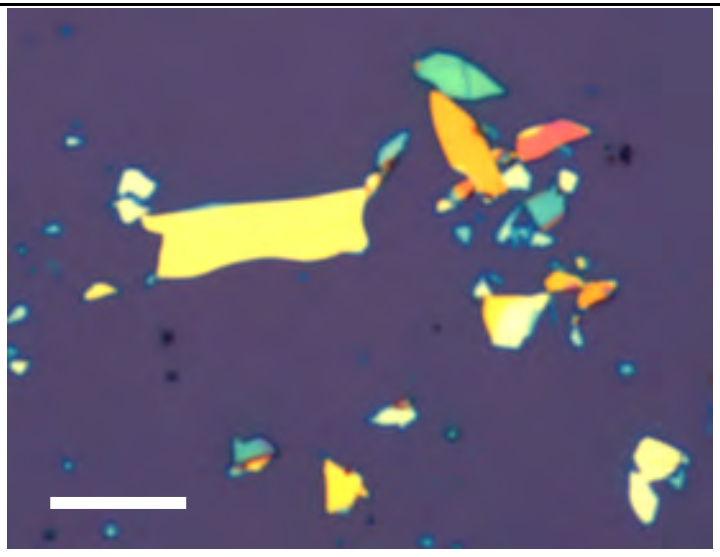 | 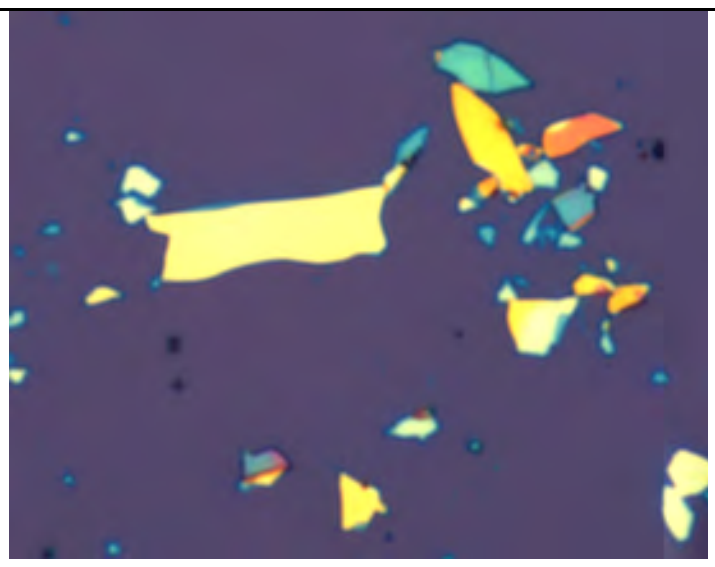 |
| ***n*-undecylamine** | **180** | 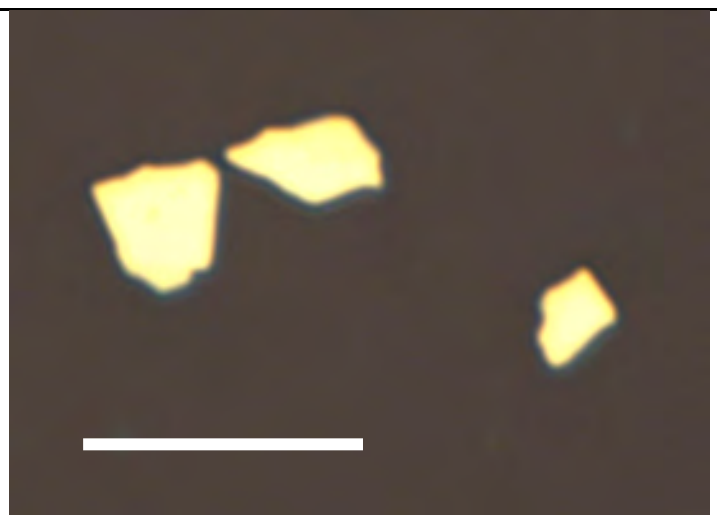 | 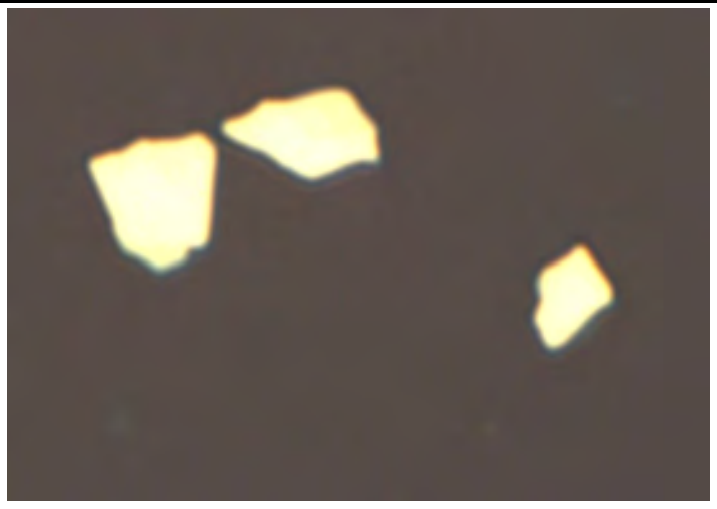 |
| **hexane-BP** <br> (not protected) | **90** | 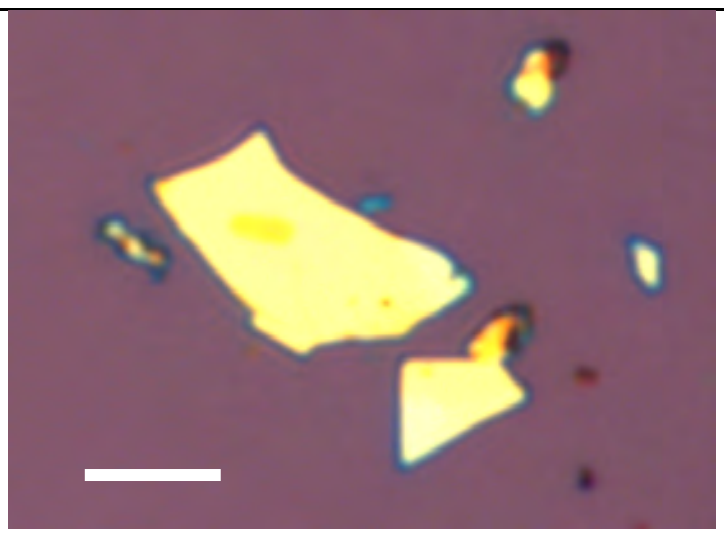 | 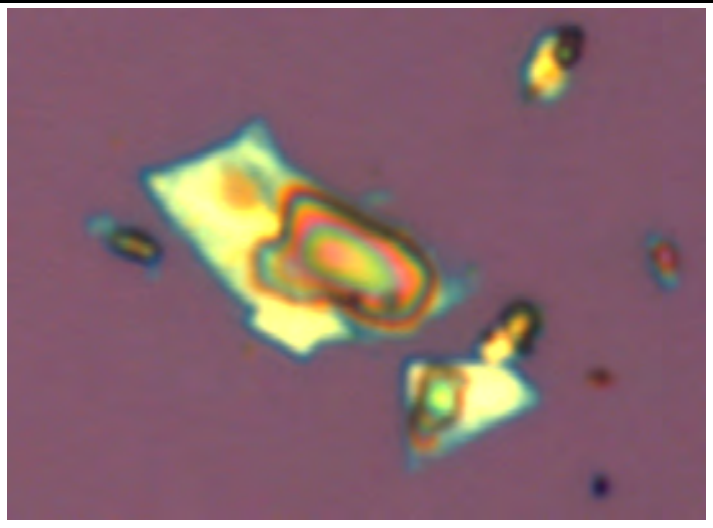 |
| **benzylamine-BP** <br> (not protected) | **120** | 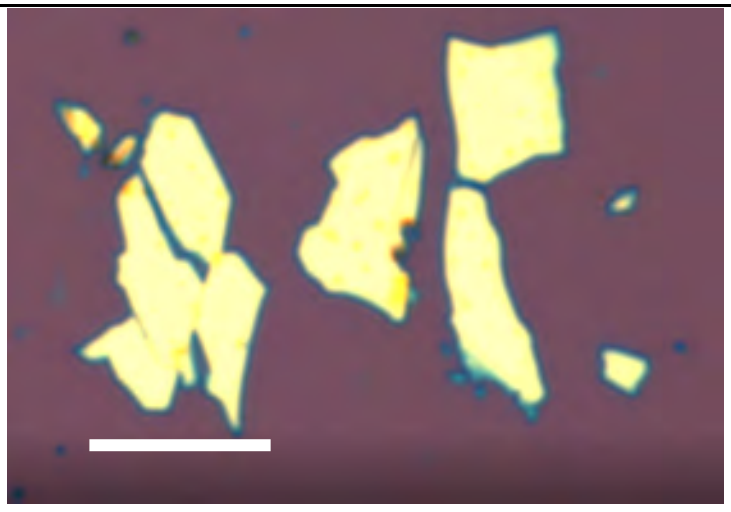 | 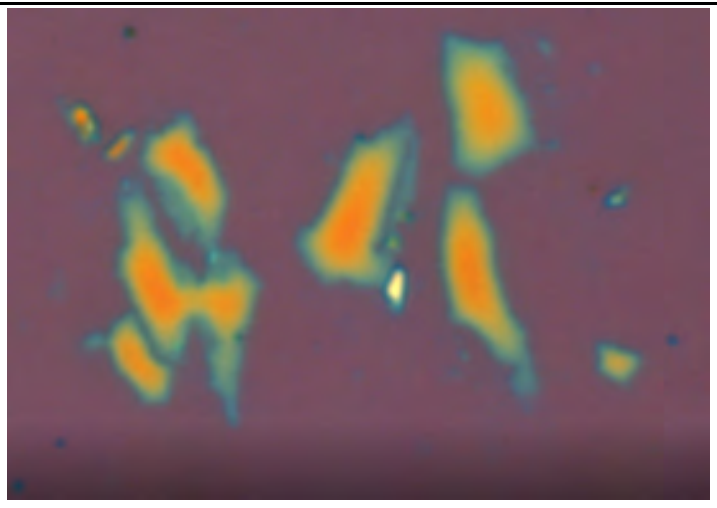 |
| | **150** | 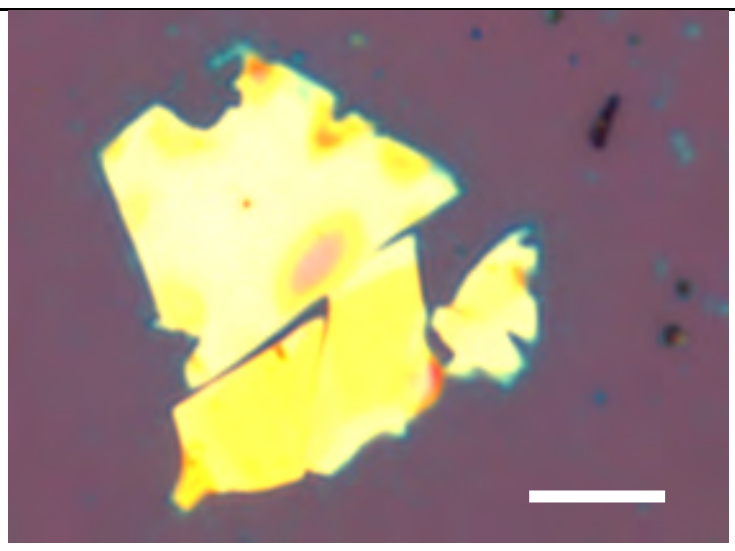 | 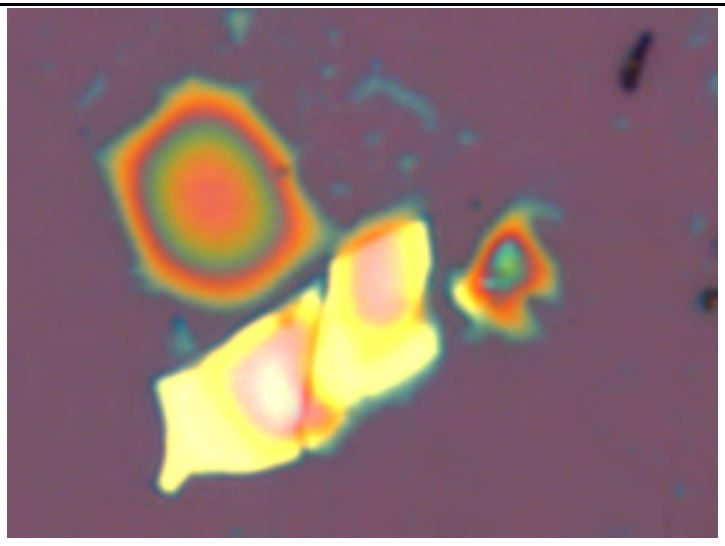 |
| | **180** | 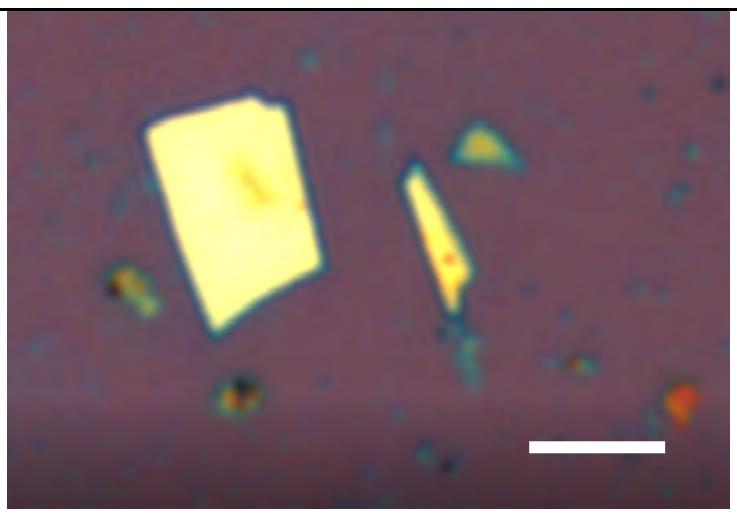 | 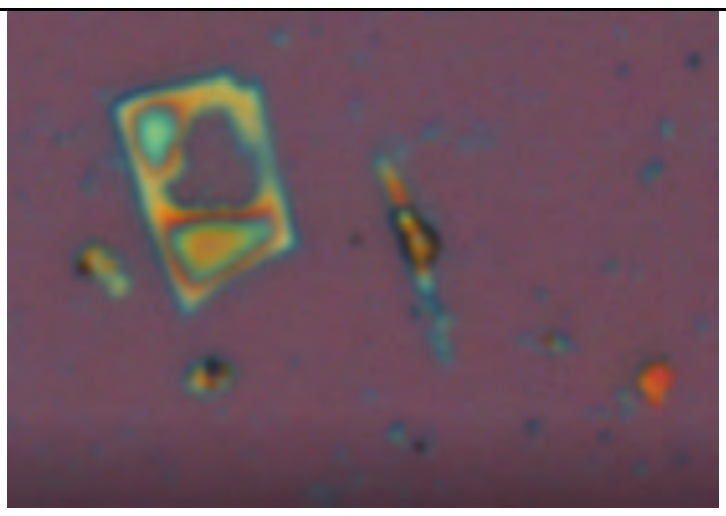 |

**Table S3. Coating parameters used for coating amines and hexane on 2D materials.**

| | **Boiling point (°C)** | **Coating temperature (°C)** | **Coating time** | **Post-coating annealing** |
|---|---|---|---|---|
| ***n*-$C_4H_9NH_2$** | ~ 78 | 90 | Two steps (1st step: 20 min heated in liquid; 2nd step: 20 min steamed in vapor) | 200 °C for 30 min in argon |
| ***n*-$C_5H_{11}NH_2$** | 105 | 110 | | |
| ***n*-$C_6H_{13}NH_2$** | 131.5 | 130 | | |
| ***n*-$C_8H_{17}NH_2$** | ~176 | 140-160 | | |
| ***n*-$C_{10}H_{21}NH_2$** | ~217 | 150-180 | | |
| ***n*-$C_{11}H_{23}NH_2$** | ~240 | 180 | | |
| **$C_6H_5CH_2NH_2$** | 185 | 180 (not protected) | | |
| ***n*-$C_6H_{14}$** | 68.7 | 80 (not protected) | | |

## 2. Characterization methods

I. Contact angle measurement

The contact angle measurement was done using a home-made zoom-in microscope.[2,3] The analysis is processed with the “contact angle” plug-in software developed by Marco Brugnara in ImageJ software. The measurement results on $SiO_2$ (figs. S1A and B) indicate no *n*-hexylamine was deposited onto $SiO_2$/Si wafer. This test also verifies the eligibility of AFM measurement on height change of 2D flakes before and after amine coating, as the flake thickness can be consistently referenced to the surface of $SiO_2$/Si wafer.

On the contrary, for $WS_2$ (figs. S1C and D) and BP (figs. S1E and F), there is an increase in the contact angle after coating of *n*-hexylamine, indicating the successful coating of amine molecules on $WS_2$ and BP. It should be noted that the contact angle only qualitatively reveals the surface wettability for these two materials, as the surface of Si wafer is not totally covered by $WS_2$ or BP flakes.

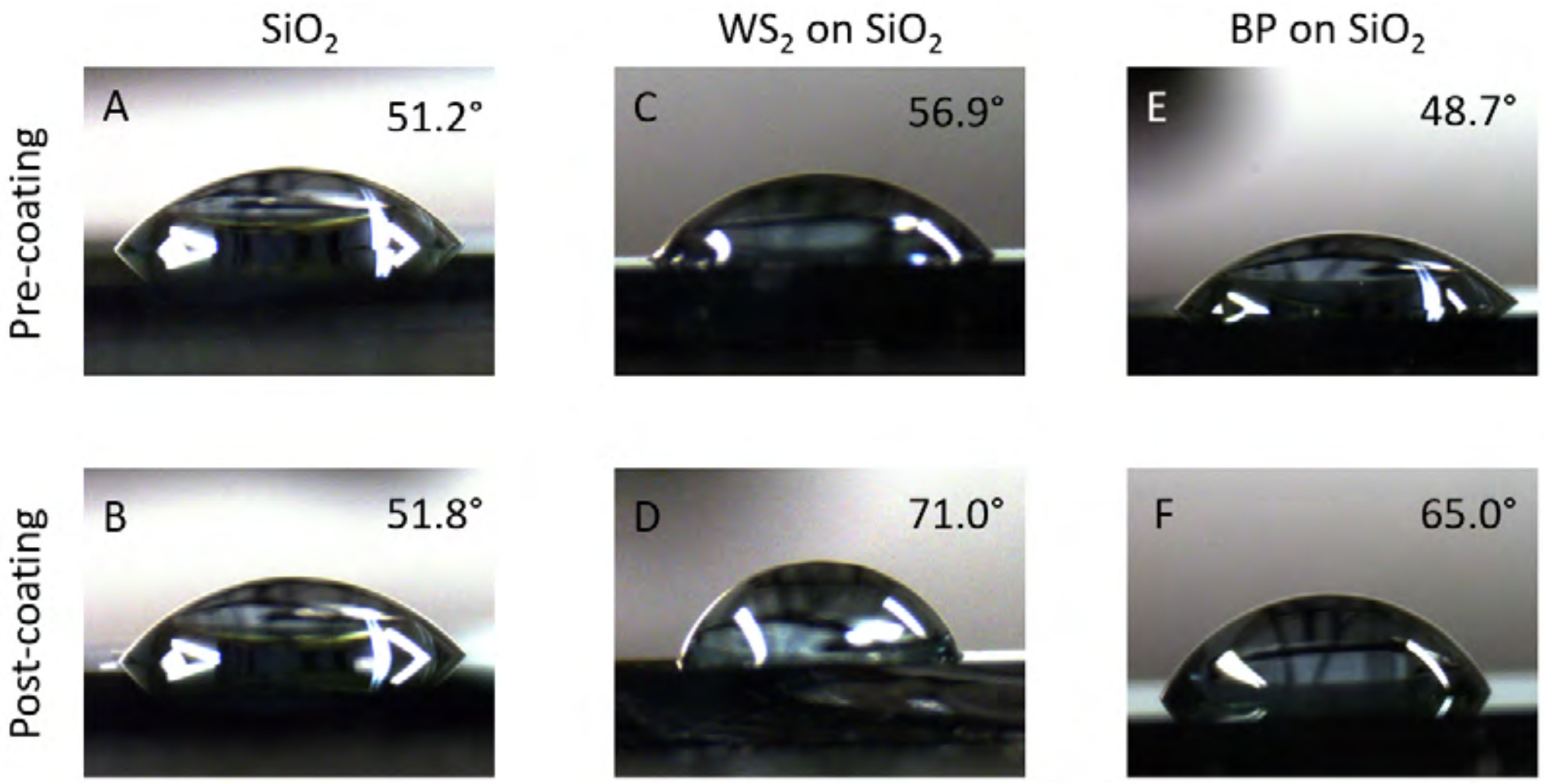

**Fig. S1. Contact angles of water droplet on different substances.** The images of a water droplet on **(A, B)** $SiO_2$, **(C, D)** $WS_2$ flakes on $SiO_2$, and **(E, F)** BP flakes on $SiO_2$ before and after *n*-hexylamine-coating process, respectively. The contact angles are marked within the images. Note that since the flakes are scattered on wafer and cover about 10% of the surface area, the angle in these image is a qualitative revelation of the hydrophobicity of the *n*-hexylamine-coated layer. The angles are averaged from ten sites on each sample.

II. Optical microscopy and Raman spectroscopy

All the 2D samples in this work were imaged using an optical microscope (Axio Imager (Carl Zeiss)). The Raman spectra of all the 2D materials were measured using a Horiba Jobin-Yvon HR800 Raman Spectrometer. The laser spot size is 1 μm in diameter and a 100× objective with an NA of 0.90 was used. The 532 nm frequency-doubled Nd:YAG excitation laser was used with the laser power on the sample set about 1 mW. 1800 lines/mm grating was used for Raman measurement. All the Raman spectra are original data with only background subtraction, and we performed the measurement under the same alignment/orientation each time.

III. X-ray photoelectron spectroscopy (XPS) and surface coverage estimation

The surface chemical analysis was carried out via a Thermo Scientific K-Alpha X-ray XPS using a monochromated Al-Kα X-ray source ($hv$ = 1486.6 eV). Each data point in the time-evolved oxidation measurement was done by opening the pump lid and letting the sample be exposed to ambient air for a certain amount of time, and then re-pumping the chamber and doing the measurement. XPS spectra of a HA-BP as coated, after 2 days, and after 46 days are shown in Fig. 2F, and the spectrum is deconvoluted by two P 2p main peaks (centered at 129.68 eV and 130.68 eV), and three oxidized phosphorous peaks (centered at 131.48 eV, 132.46 eV, and 134.08 eV).[4] The phosphorous percentage is calculated by analyzing the composition of these peaks, and fitted by an exponential growth curve in Fig. 2G. The curve has the form $\eta = A - B(1 - e^{-\frac{t}{\tau}})$, where $\eta$ is the percentage of oxidized phosphorous, $A$ and $B$ are fitting parameters, and $\tau$ is the time constant. According to the fitting curves of BP and HA-BP in Fig. 2G, we find that $\tau_{HA-BP} \sim 32\tau_{BP}$. From the diagram, the whole curve of HA-BP clearly cannot be fitted by a single exponential function. A second curve starting at the ~100 hours with a much larger time constant is shown, and this could be ascribed to the saturation of the passivation layer, where the passivation layer itself could drastically slow down the further oxidation process. However, it is seen that after 48 hours, the oxidation of HA-BP clearly slows down, which is fitted by the dashed curve. This slow-down may be due to the saturation of surface oxidation of phosphorous and the phosphorous oxide forms a passivation layer that further protects the phosphorous beneath. According the fitting curves of BP and HA-BP, we find that $\tau_{HA-BP} \sim 32\tau_{BP} (< 50h)$ at the starting period and $\tau_{HA-BP} \sim 238\tau_{BP}$ (>

$50h$). The change of tunneling rate of oxygen should follow an Arrhenius relation as $\exp(-\Delta E/k_B T) < 1/32$, thus $\Delta E > 0.09$ eV, which indicates that the coverage should be above 66.7% (whose tunneling barrier is 0 eV for $O_2$) by combining the calculated results in the work.

## IV. Atomic force microscopy (AFM)

For BP measurement, tapping mode atomic force microscopy is done using a Nanosurf Flex-Axiom AFM enclosed in a glove box with an inert environment containing <0.1 ppm $O_2$ and <0.1 ppm $H_2O$ to avoid corrosion. For $WSe_2$ measurement, the measurements were performed on Dimensional Icon system (Bruker, USA) in tapping mode with ultrasharp probes (OMCL-AC160TS, Olympus, Japan). The resonance frequency and spring constant for the probes are about 285 kHz and 26N/m, respectively. The thickness and roughness were obtained by analyzing the AFM height image with Scanning Probe Image Processor software (Image Metrology Aps, Denmark).

## 3. First-principles calculation

### I. Calculation method details

Calculations are performed using Vienna ab initio simulation package (VASP)[5] with the generalized gradient approximation of Perdew-Burke-Ernzerhof (PBE)[6] for the exchange-correlation potential and a projector augmented wave (PAW) method.[7] The DFT-D3 method[8] was adopted to account for the van der Waals interactions. Supercells containing a vacuum spacing larger than 15 Å were used to model the BP and $WS_2$ surfaces. The kinetic energy cutoff for plane wave functions is set to 500 eV and the energy convergence threshold is set as $10^{-4}$ eV. The Monkhorst-Pack k-mesh[9] of 5×5×1 is employed to sample the irreducible Brillouin zone. The atoms were fully optimized and the maximum force on each atom is less than 0.01 eV/Å. Bader's charge analysis is done for analyzing the charge distribution after proton transfer.[10,11] For hexylamine-BP system, two BP bilayers were used for the model with the bottom bilayer was fixed during optimization (not shown in following figures); while for hexylamine-$WS_2$ system, only a single layer of $WS_2$ was included in the model.

### II. Adsorption energies of different configurations for hexylamine−BP system

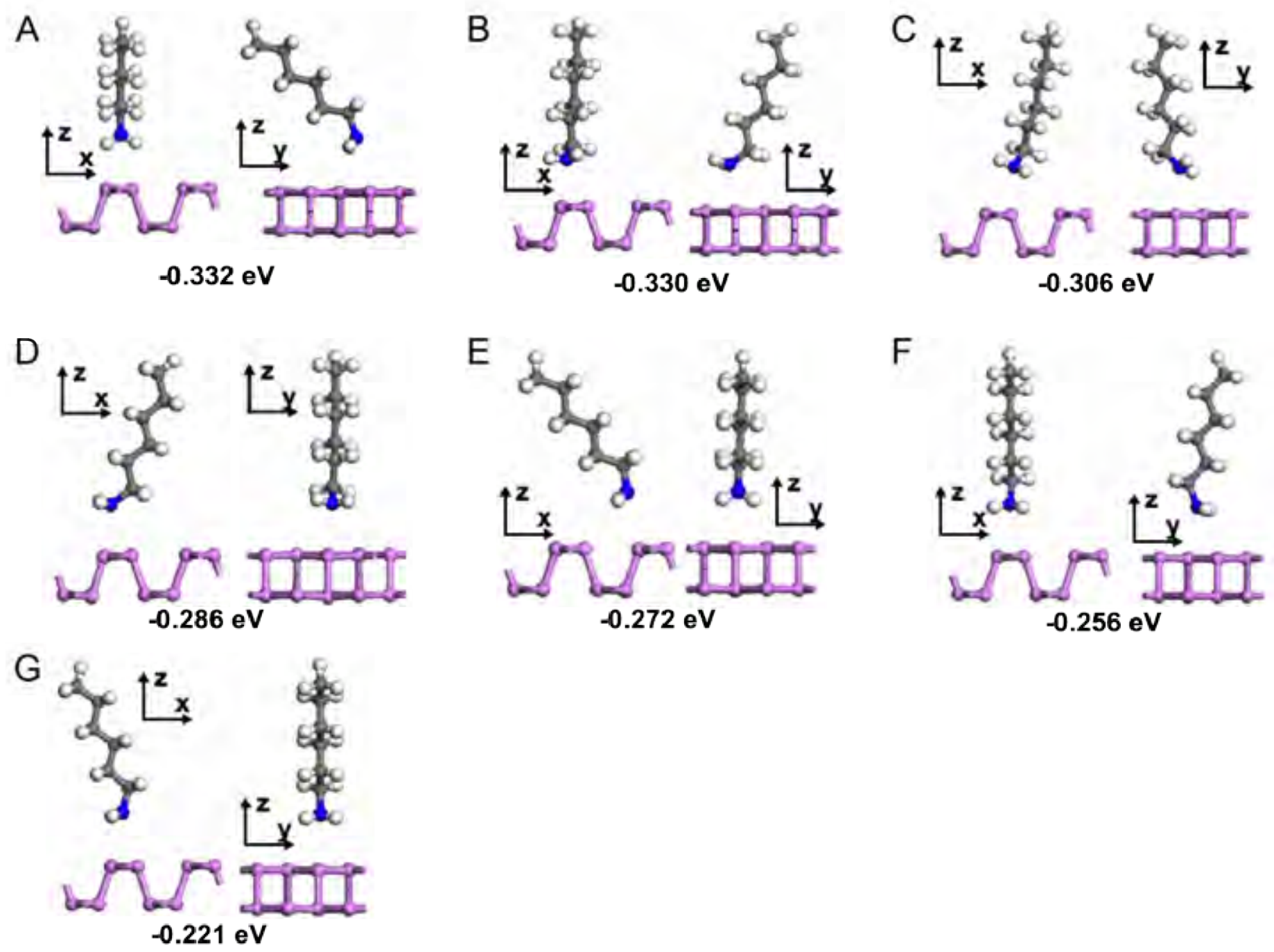

**Fig. S2.** Seven different configurations for direct adsorption of *n*-hexylamine on BP. For each configuration, the front view and side view are shown, and the adsorption energy is marked under each configuration. The largest adsorption energy is only 0.33 eV.

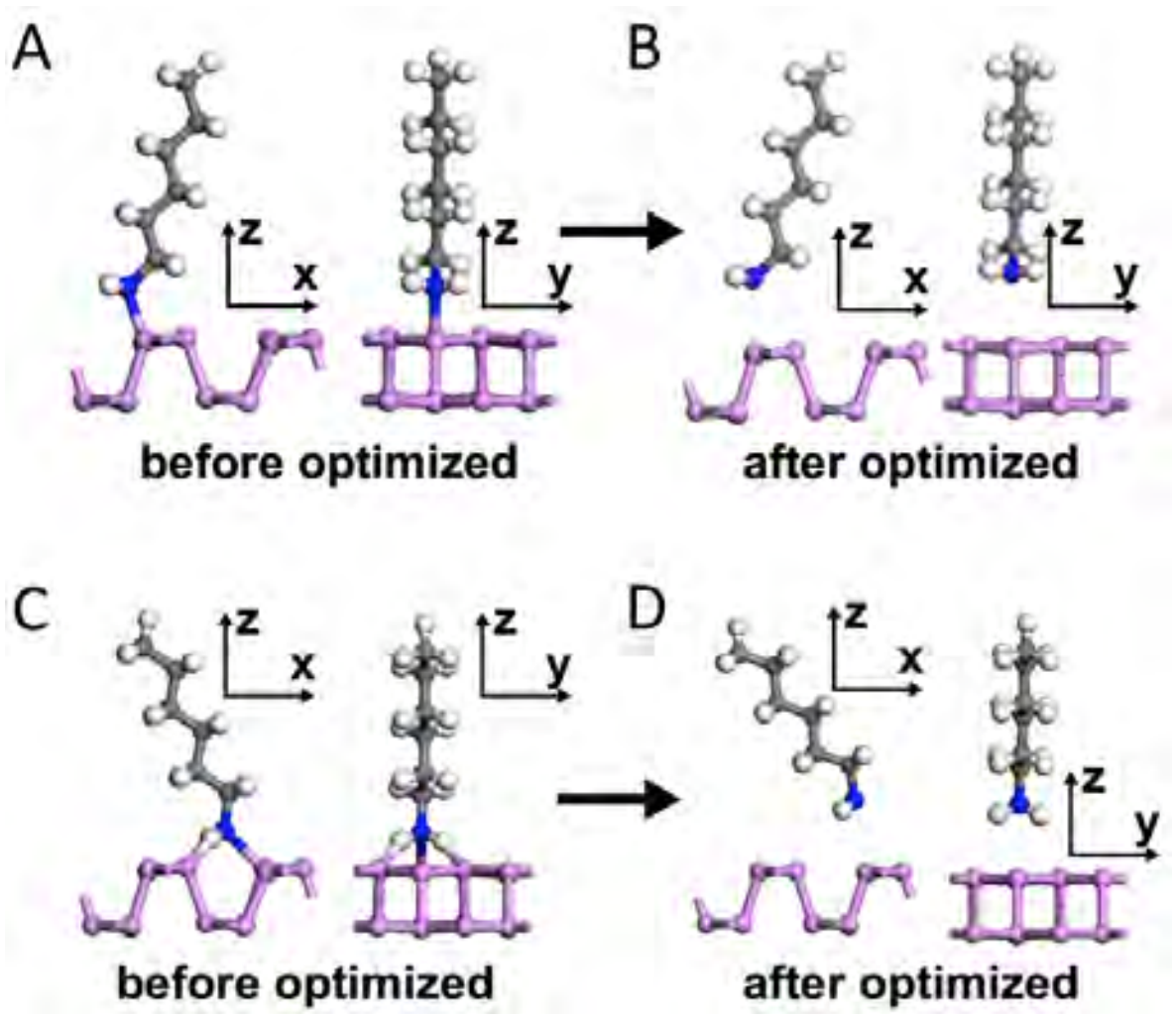

**Fig. S3.** **(A and C)** For direct adsorption of *n*-hexylamine on BP, two special initial configurations have been tested, in which *n*-hexylamines are chemically bonded with BP. **(B and D)** After optimization, *n*-hexylamines in these two configurations have been repelled by BP, indicating that direct chemical bonding between *n*-hexylamines and BP is not possible.

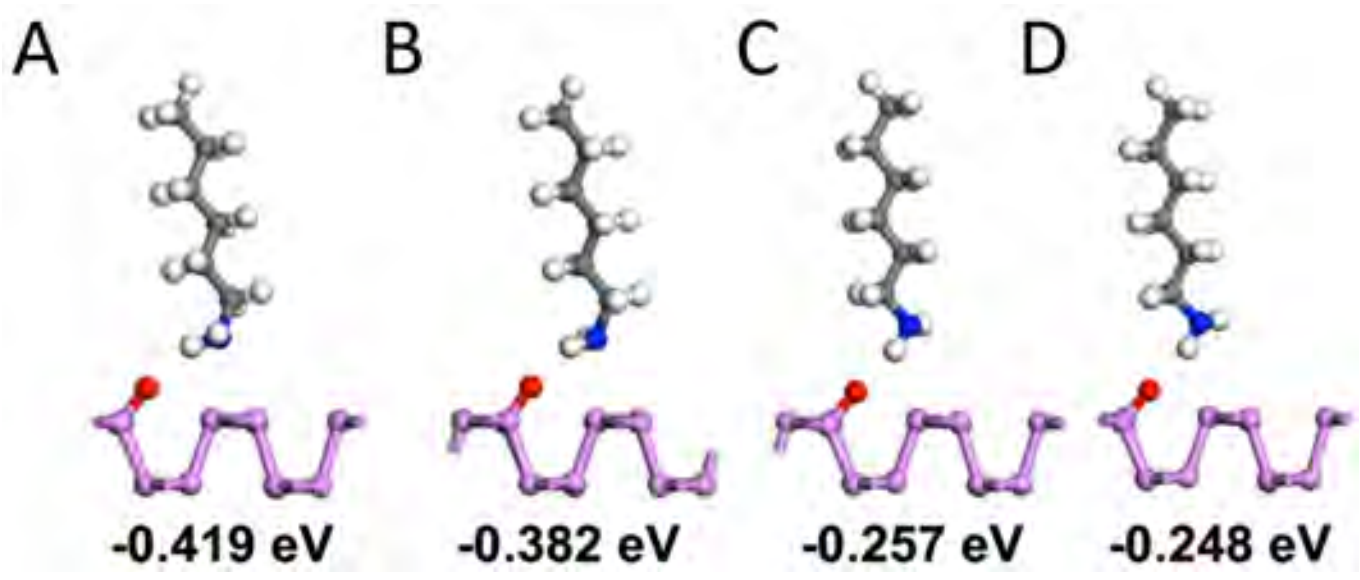

**Fig. S4.** Four different configurations for adsorption of *n*-hexylamine on oxidized BP. The largest adsorption energy is 0.419 eV, indicating that oxidization of BP can enhance adsorption between BP and *n*-hexylamine.

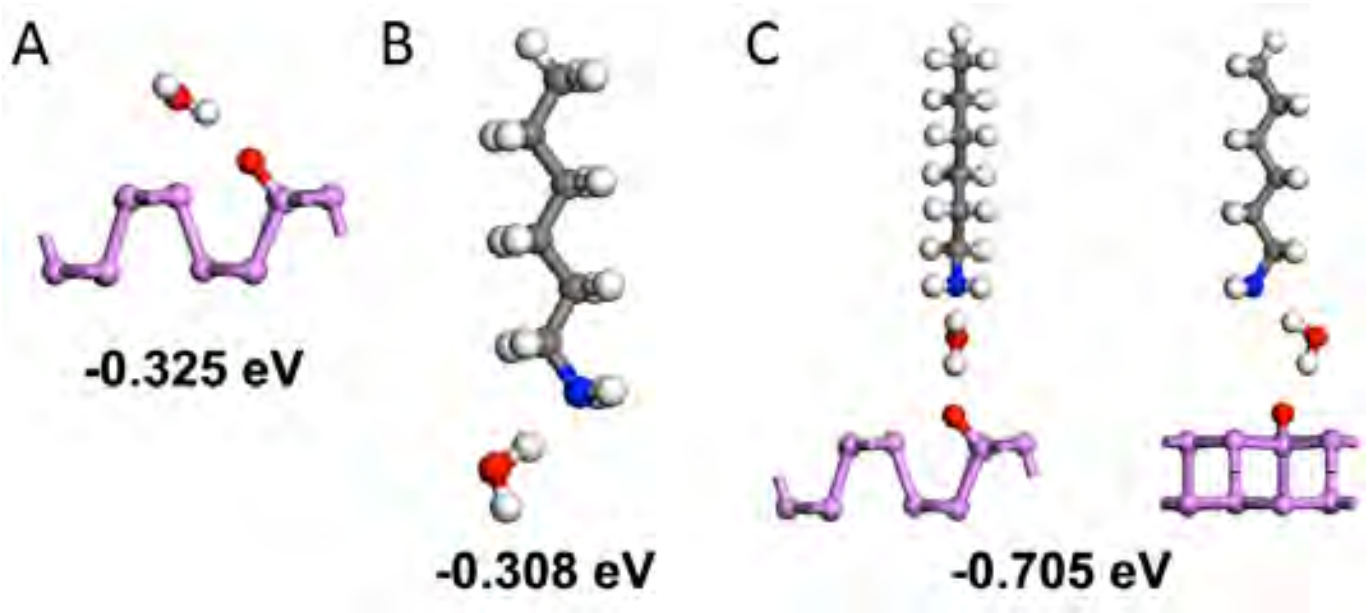

**Fig. S5.** **(A)** The lowest energy configuration for adsorption of $H_2O$ on oxidized BP. **(B)** The lowest energy configuration for adsorption between *n*-hexylamine and $H_2O$. **(C)** The lowest energy configuration for $H_2O$-mediated adsorption, where the adsorption energy consists of the two binding energies between $H_2O$ molecule and oxidized BP and between $H_2O$ and *n*-hexylamine.

III. Surface coverage for hexylamine−BP system

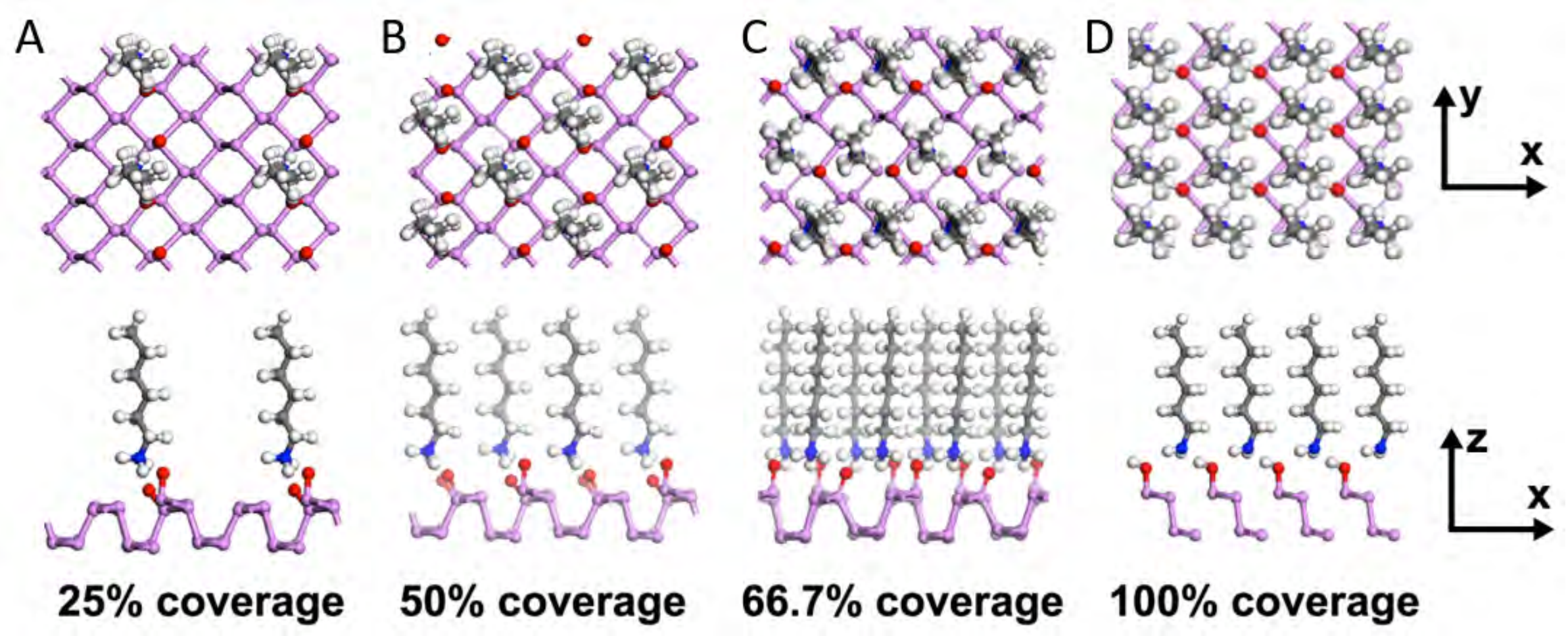

**Fig. S6.** The models of **(A)** 25% coverage (binding energy 1.07 eV/*n*-hexylamine), **(B)** 50% coverage (binding energy 1.2 eV/*n*-hexylamine), **(C)** 66.7% coverage (binding energy 1.18 eV/*n*-hexylamine), and **(D)** 100% coverage (binding energy 0.58 eV/*n*-hexylamine), used for calculating the penetration energy barrier of $H_2O$ molecules and $O_2$ molecules. Note that in 100% coverage, the footprint of *n*-hexylamine molecules completely covers the BP, so that no denser coverage could be further achieved. The first row are top views and the bottom row are side views of corresponding configurations.

IV. Energy barrier calculations during kinetic transport for hexylamine−BP system

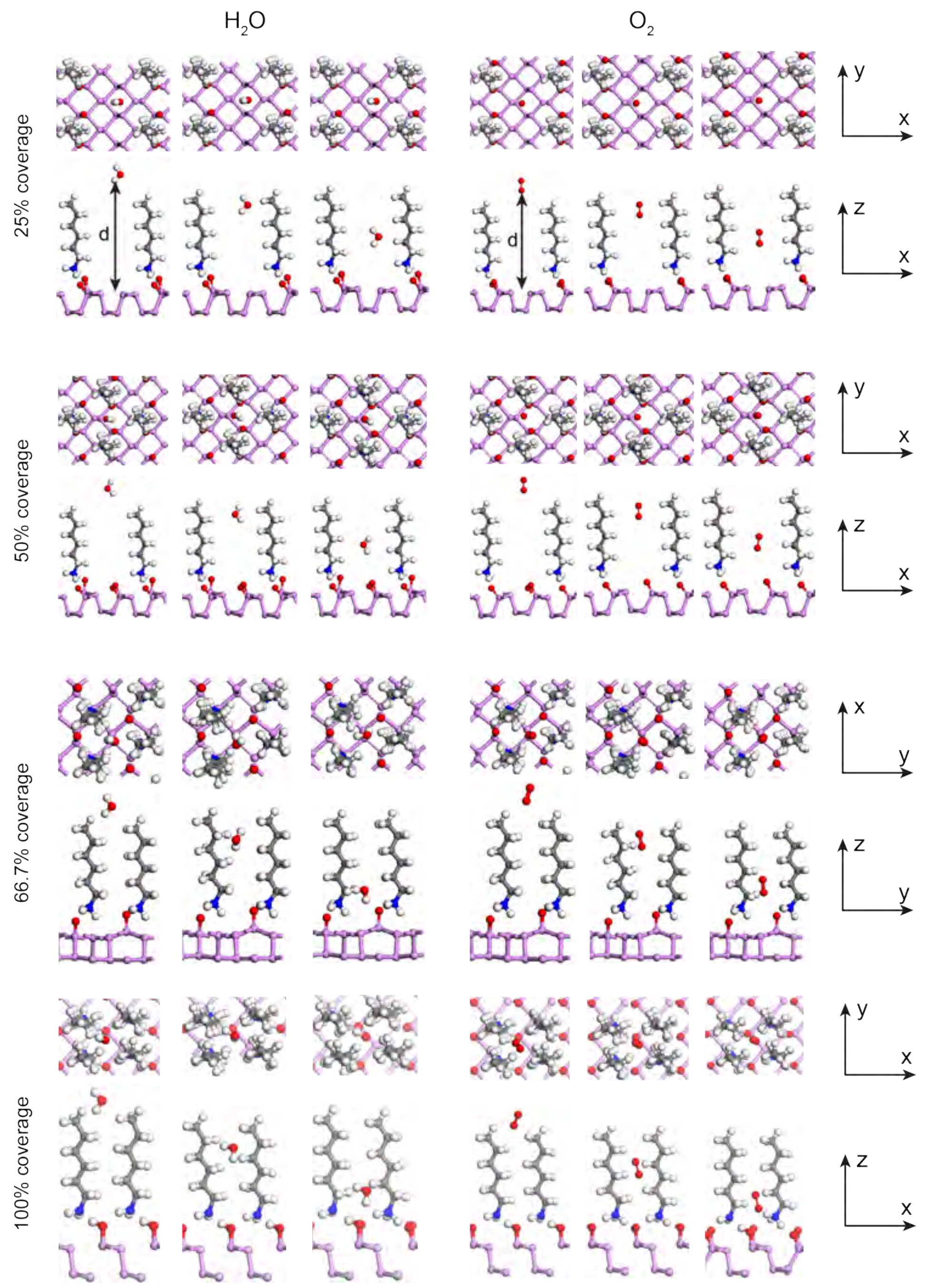

**Fig. S7.** The structures of HA-BP used for calculating $H_2O$ and $O_2$ molecules penetrating through the *n*-hexylamine coating. For each coverage and penetrating molecule type, three different locations are shown (far, middle, and close to the BP surface respectively), and each location is shown in two perspectives which are top view and side view. The distance *d* is defined in the first two figures of $H_2O$ and $O_2$ in 25% coverage.

## 4. Photodetector device

### I. BP photodetector fabrication

First, the pattern with Ti(15 nm)/Au(150 nm) for marking was fabricated onto $SiO_2$(190 nm)/Si substrate by photolithography. These Ti/Au patterns help to locate the BP flakes that will be exfoliated onto the $SiO_2$/Si substrate subsequently. Before BP exfoliation, the $SiO_2$/Si substrate was annealed/dried in air for 30 minutes at > 200°C to remove humidity from air and then transferred into a glove box. Exfoliation of BP flakes onto $SiO_2$/Si was performed in the glove box. After exfoliation, the metallic electrode pads of Ti(15 nm)/Au(200 nm), fabricated onto BP flakes sitting on $SiO_2$/Si substrate were defined by electron-beam lithography (EBL) using PMMA as the resist mask, where the active channel width and length are kept comparable for both *n*-hexylamine-protected and the control BP detectors. After liftoff of PMMA, the BP photodetector fabrication was completed for the BP control sample. To prepare the *n*-hexylamine-protected BP detector, the amine molecules were coated onto BP devices with the aforementioned method prior to characterization.

### II. Channel BP thickness measurement with AFM

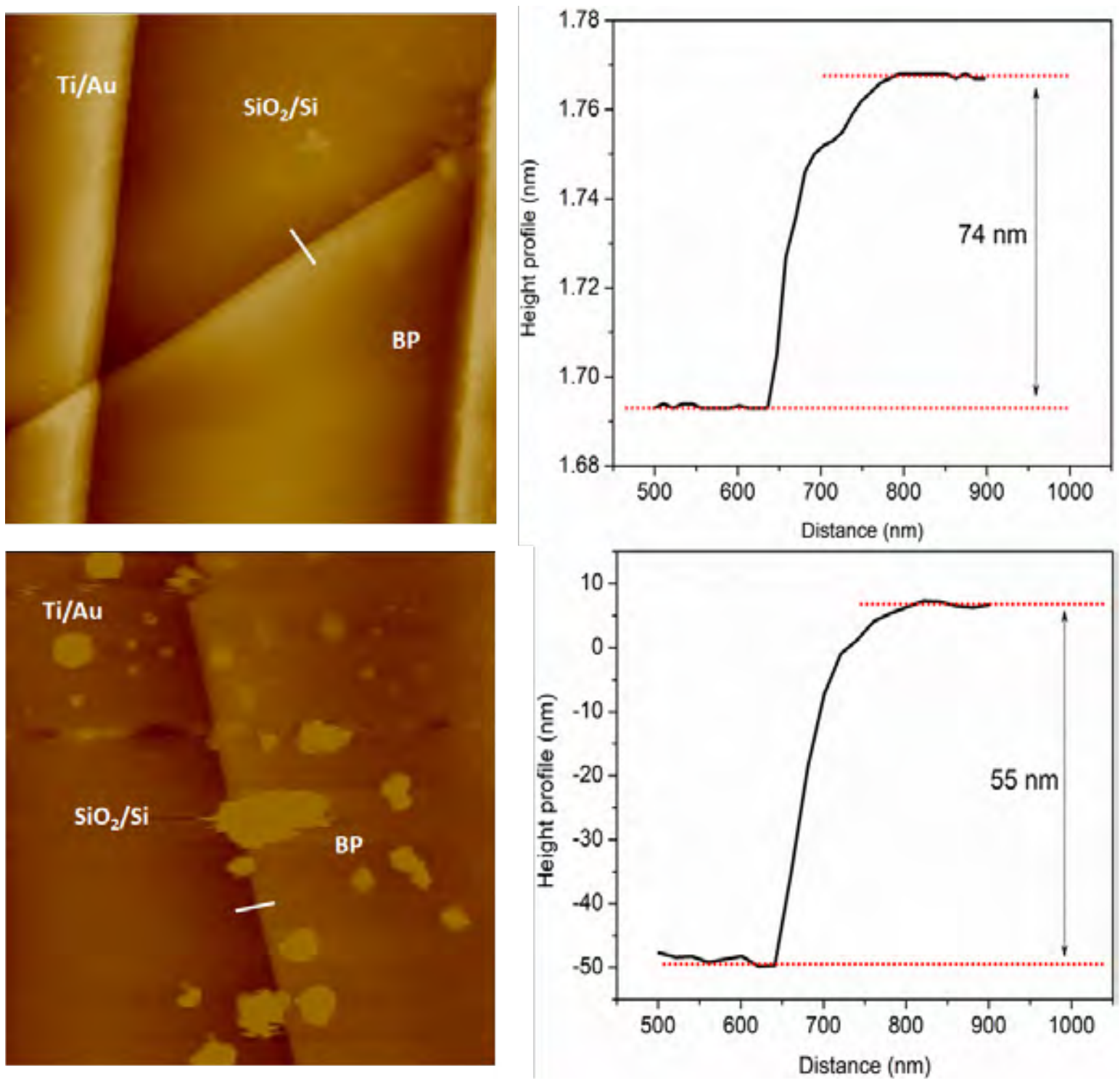

**Fig. S8.** Height profile for the channel of BP flakes in pure BP control device with a thickness of 74 nm (upper panel) and HA-BP device with a thickness of 55 nm (lower panel). The particles on the HA-BP device are residues from $H_2O_2$ solution after treatment of the device with $H_2O_2$ etchant. The dimension of the two AFM topographic images is 5μm×5μm.

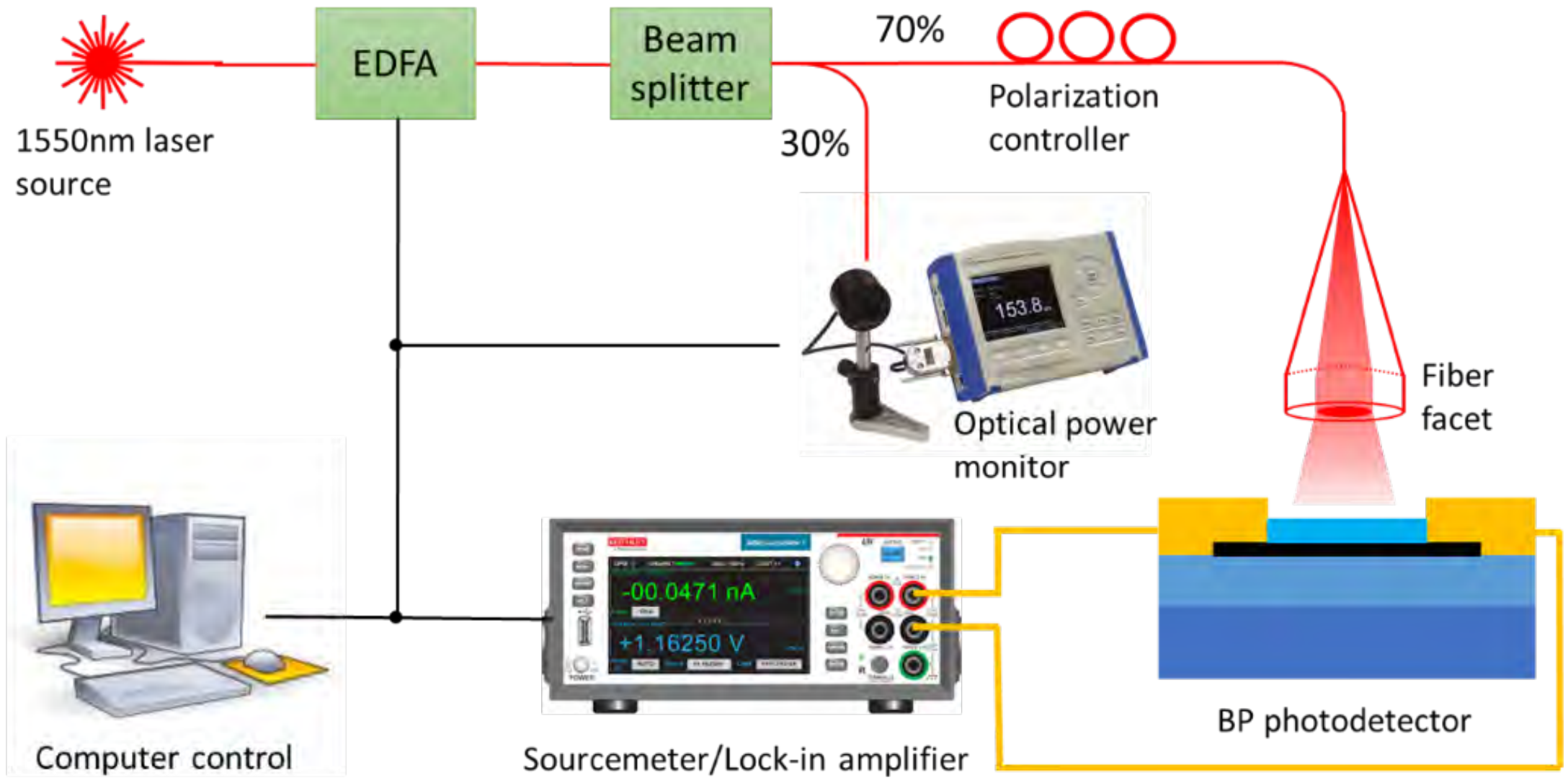

**Fig. S9.** Schematic of the photodetector measurement setup.

III. Measurement setup for characterizing black phosphorus photodetectors

Figure S9 shows a schematic diagram of the system setup. An all-fiber system was used to deliver probing light to the photodetector. Light from a 1550 nm tunable external cavity laser was coupled into an optical fiber and amplified by an Erbium Doped Fiber Amplifier (EDFA, Amonics). The EDFA output was split into two beams: 30% of the light was monitored by an optical power meter (Newport 1918A) to record the light intensity in real-time during the experiment. 70% of the light passed through a fiber with a cleaved facet and incident upon the black phosphorus photodetector at a fixed incident angle of 15 degrees. Due to the highly anisotropic optical and electrical properties of black phosphorus, an inline polarization controller was used to adjust the polarization of light to maximize the photocurrent. A source meter (Keithley SMU2450) was used to measure the photocurrent generated by the photodetector at zero bias while laser was working on continuous wave mode. To assess the photoresponsivity under bias condition, the laser light was modulated at 30 kHz and a trans-impedance amplifier (SRS SR570) was used to apply the bias. The amplified photocurrent was recorded using a lock-in amplifier (SRS SR844). We use finite-difference time-domain (FDTD) simulations to quantify the optical power absorbed by the photodetector, shown in Figure S19. The laser light from the fiber facet is modeled as a Gaussian beam with a beam waist of about 5.5 μm according to our simulations. A power monitor was used to evaluate the power flux passing though the photodetector. The model indicates that 23% of light exiting from the fiber was captured within the active area of the photodetector. Detector responsivity was calculated by normalizing the measured photocurrent by the incident optical power on the photodetector.

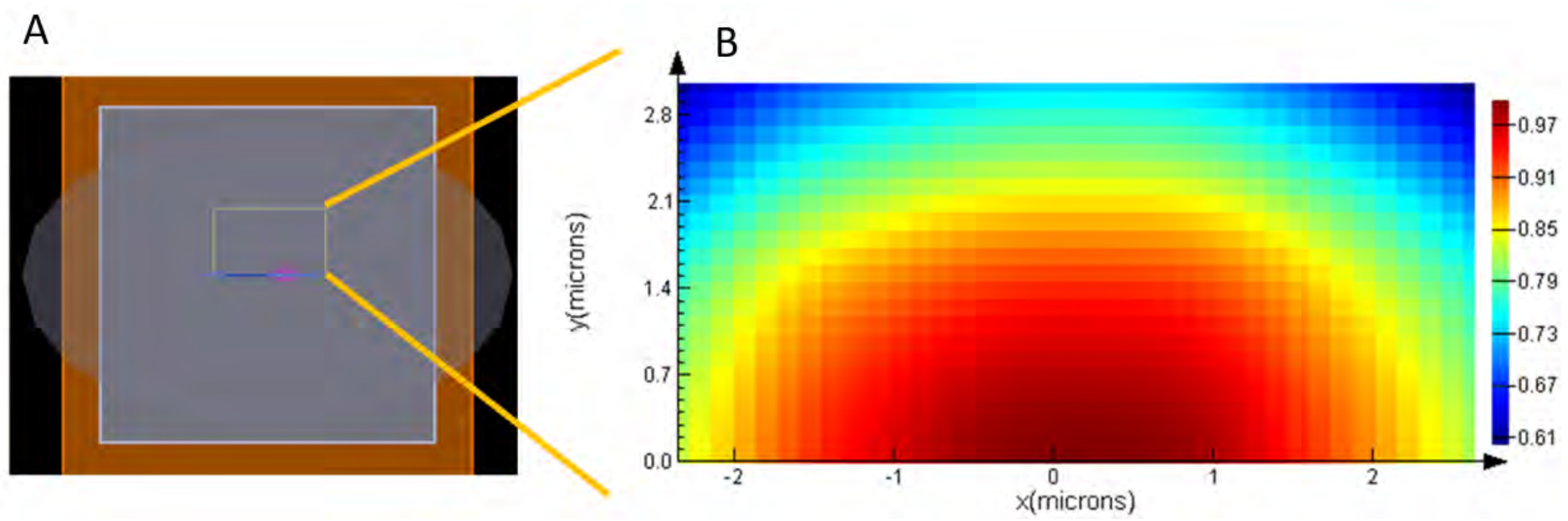

**Fig. S10.** **(A)** yellow rectangle showed the relative position between the active region of a 5 μm by 3 μm size photodetector and the center of incident laser light at optimized coupling condition. **(B)** electromagnetic field distribution on the photodetector.

IV. Photothermoelectric photocurrent generation in BP photodetectors

We experimentally investigated the photocurrent generation mechanism in the BP detectors by mapping the photocurrent as a function of the incident beam location. As shown in Figure S14A, the fiber light source, mounted on a linear translational motion stage, was traversed across the detector active area while the photocurrent was monitored. Figure S11B and 11C plot the measurement results obtained on black phosphorus detectors with and without inhibitor protection, respectively. In both types of devices, photocurrent reaches maximum when the illumination spot is close to the metal junctions and passes through zero close to the center of the device. This behavior is an unequivocal signature of photothermoelectric response.[12,13]

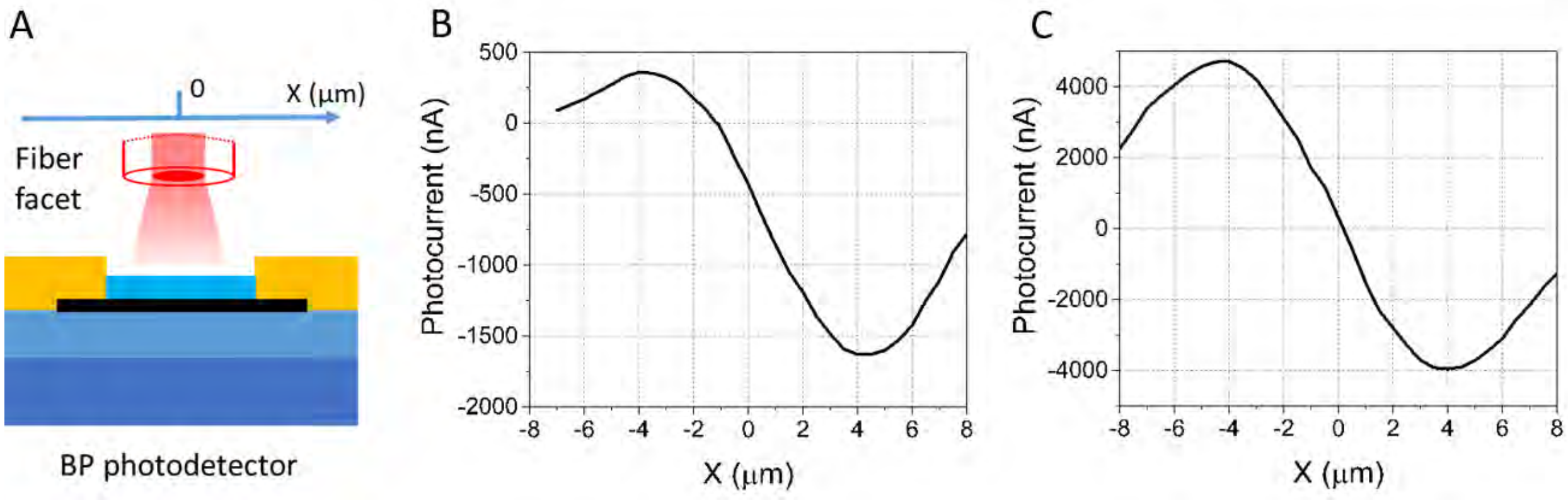

**Fig. S11.** **(A)** Illustration of photocurrent scanning set-up and photocurrent generation cross-section along the BP photodetector **(B)** with and **(C)** without *n*-hexylamine coating at zero bias condition.

## 5. Removability testing

I. *n*-alkylamine/BP

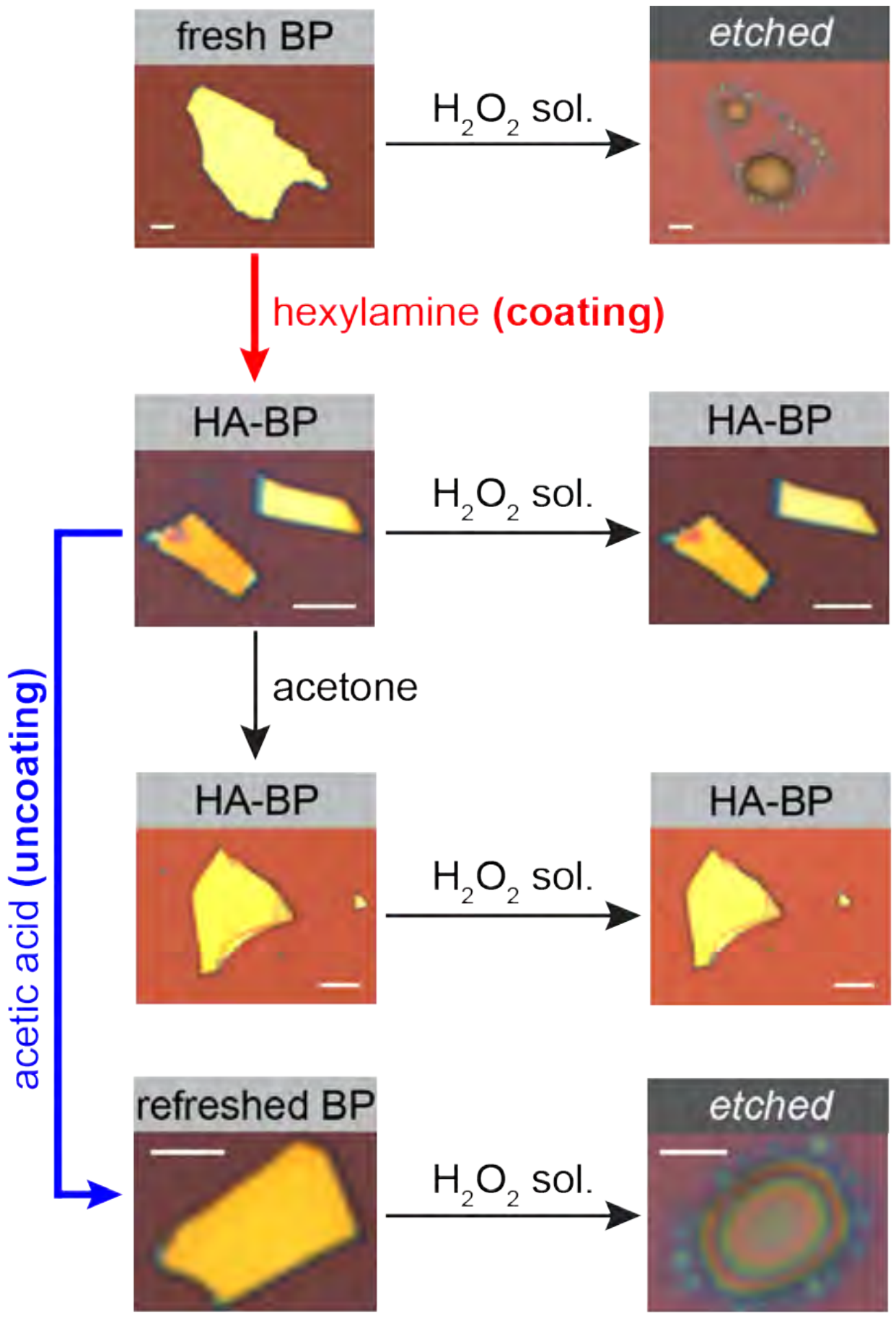

**Fig. S12. Removability of *n*-hexylamine coating on BP by organic acid.** First row, the fresh BP as a starting point, is etched by $H_2O_2$ solution (30 wt% in $H_2O$). After coating *n*-hexylamine on fresh BP, HA-BP is resistant to $H_2O_2$ etching, as shown in the second row. In the third row, the HA-BP sample is treated with acetone for 20 mins, but still resistant to $H_2O_2$ etching; while by applying glacial acetic acid or acetone/HCl (acetone : HCl (37 wt% in $H_2O$) = 1:1 in volume ratio) solution on the HA-BP for 20 mins at room temperature, the *n*-hexylamine coating is removed, and the BP can be again etched by $H_2O_2$ solution, as shown in the last row. Red arrow indicates the coating process, and blue arrow the uncoating process. The scale bars are 10 μm.

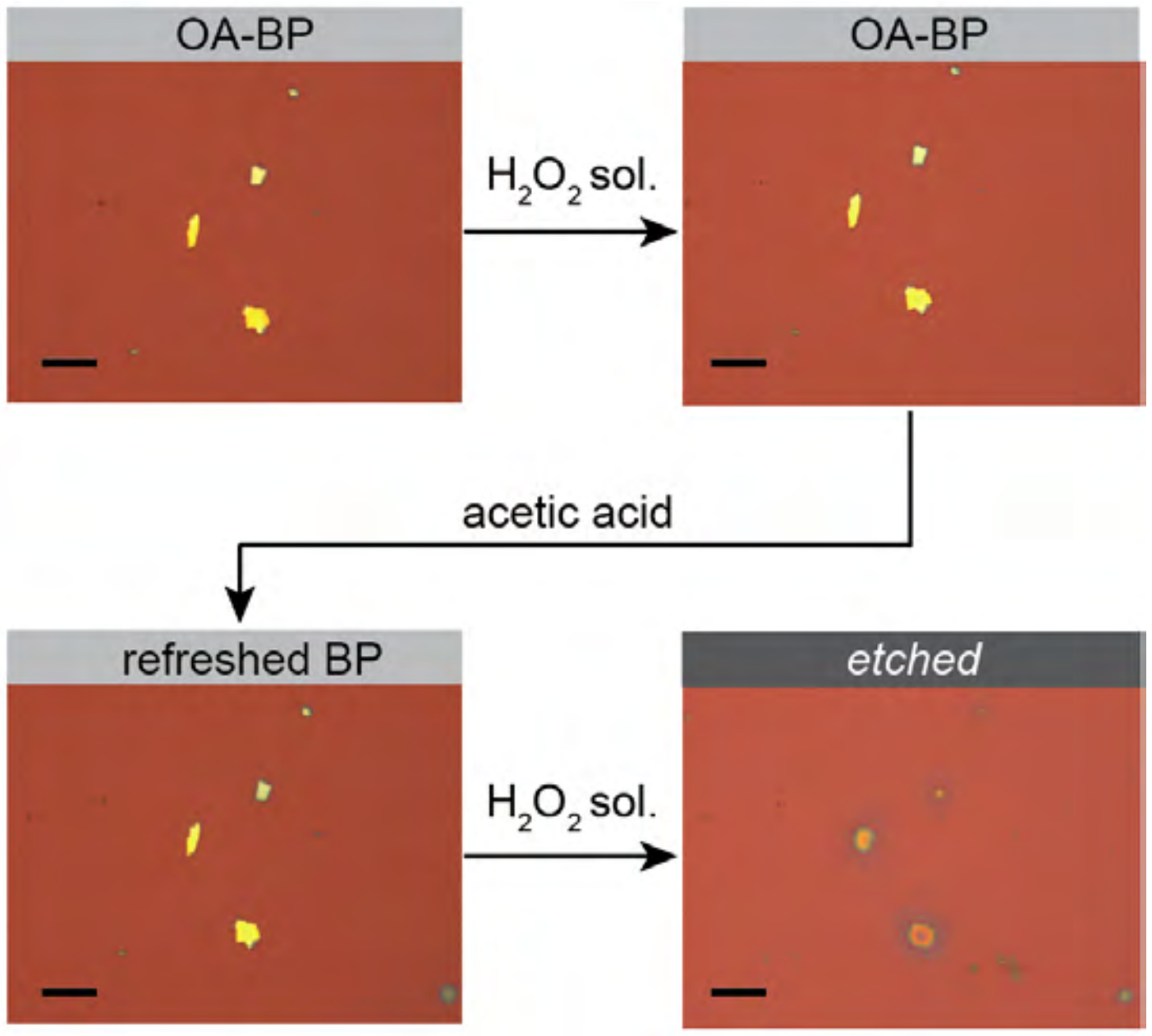

**Fig. S13**. **Removable *n*-octylamine coating on BP.** First row, the BP has been coated with octylamine and dipping inside hydrogen peroxide without etching demonstrates its successful coating. Bottom row, the coating is removed in glacial acetic acid to refresh BP surface, and then the BP is easily etched by hydrogen peroxide. The scale bars are 20 μm.

II. $n$-hexylamine/$WSe_2$

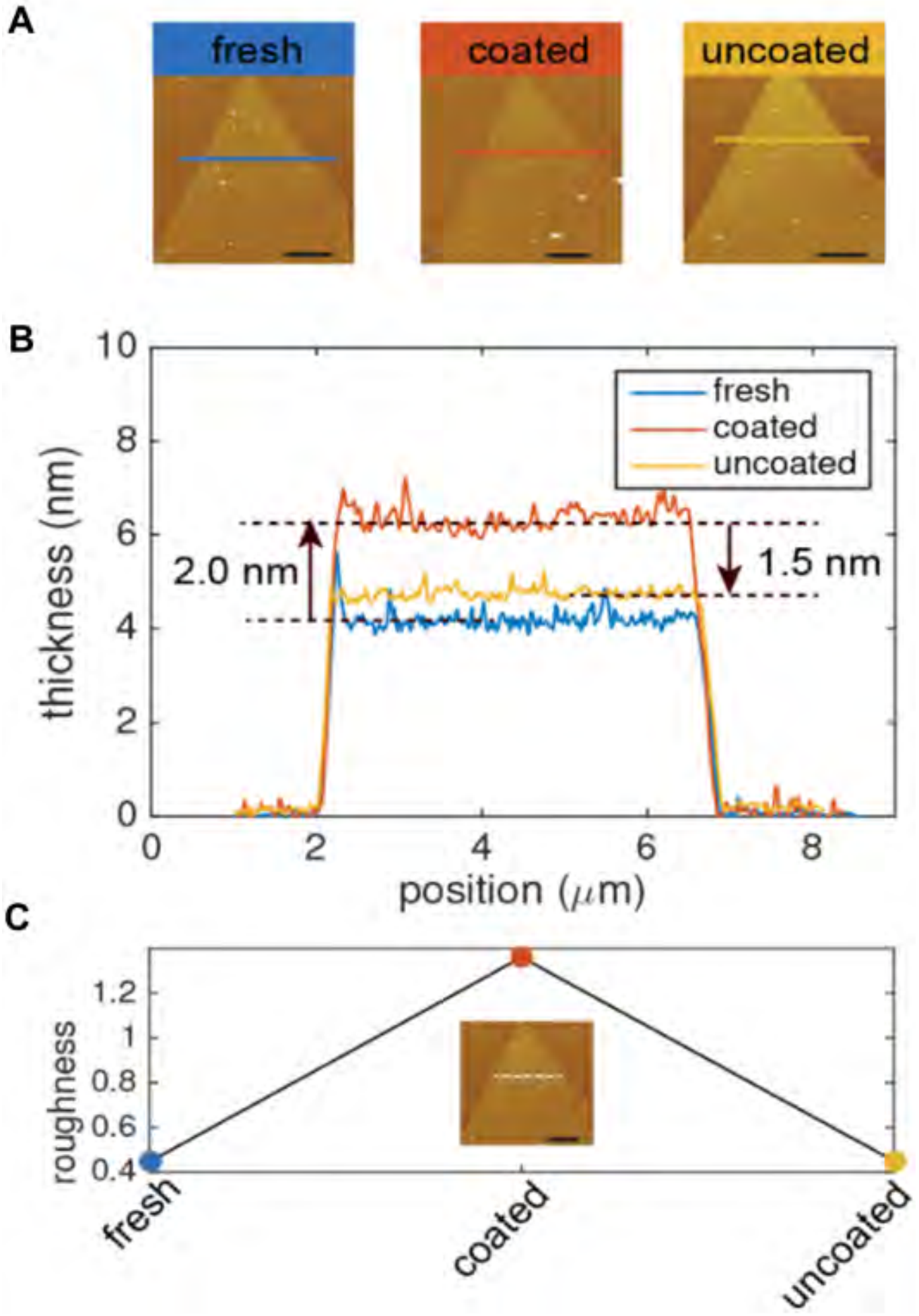

**Fig. S14. Removable *n*-hexylamine coating on $WSe_2$. A,** The AFM images of the same $WSe_2$ flake at three stages: as-exfoliated (marked as "fresh", blue color-coded), after coating (marked as "coated", scarlet color-coded), and after removing the coating (marked as "uncoated", golden color-coded). **B,** The AFM characterization of the thickness of the same $WSe_2$ flake during three stages, with the locations of height profile marked by solid lines in **A** (Note: the same color-coding scheme was used in a-c for convenient comparison). **C**, The one-dimensional surface roughness from this $WSe_2$ flake during the three stages. The roughness is measured at the same location marked by white dashed line in the inset. The scale bars are 2 μm in **B** and **C**.

As the example for transition metal dichalcogenides (TMDs), $WSe_2$ was used to carry out a thorough removability testing by monitoring the same $WSe_2$ flake's height, roughness and Raman characteristics during the three steps in treating the flake – fresh uncoated, $n$-hexylamine coated, and coating removed (fig. S14). $WSe_2$ flakes were exfoliated on $SiO_2$/Si substrate by mechanically exfoliation. The thickness of $WSe_2$ flakes were measured by AFM at different states (before coating hexylamine, after coating hexylamine and after removing hexylamine). To coat hexylamine, the samples were immersed in hexylamine solvent, and heat to 130 °C for 20 min. Then the sample was washed in hexane, and dry by nitrogen. To remove the hexylamine, the samples were immersed in warm acetic acid (about 50 °C) for one hour.

As shown in fig. S14c, the surface roughness of the flake is totally restored after removing the coating, and the coated *n*-hexylamine layer is 1.5 nm thick. Raman signal does not change before coating and after coating removed fig. S15. Note here, the contribution in Raman from the ultra-thin coated *n*-hexylamine layer is not observable, and hence no obvious change in Raman before and after coating for flake. However, the more significant information from Raman is that the flake property was retained from coating till coating removed, thus further consistently demonstrating our developed alkylamine monolayer coating is strong but removable.

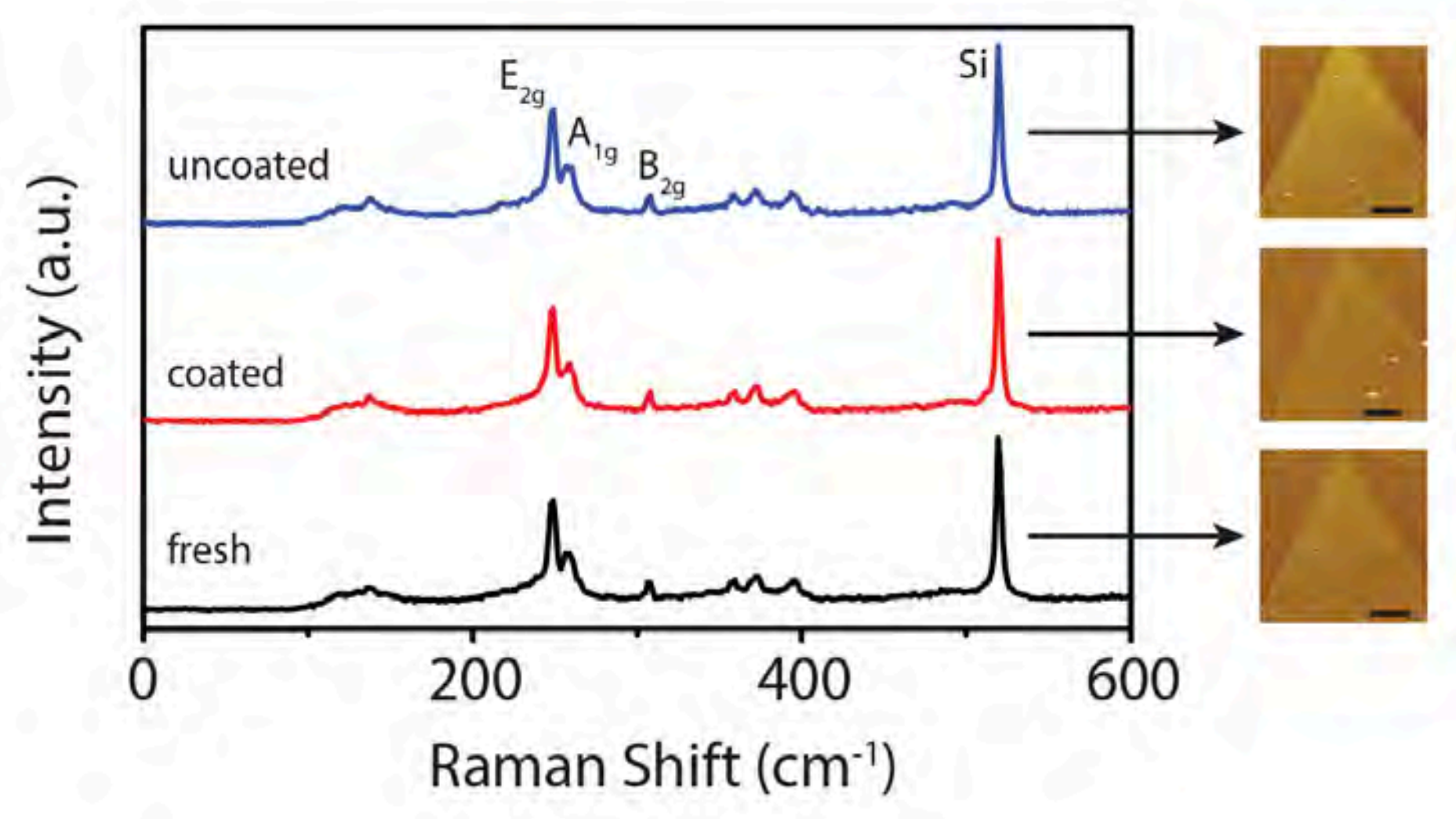

**Fig. S15. Raman properties from $WSe_2$ during removable testing**. Raman spectra taken at the same location during the three steps of removability test on *n*-hexylamine coating on the same $WSe_2$.

## 6. Protection techniques comparison

**Table S4. Comparison between existing protection techniques.** Factors such as coating layer thickness, resistive property, and techniques used are compared.

| | Thickness | Passivation against: | Susceptible to: | Fabrication method | Limitations |
|---|---|---|---|---|---|
| *n*-hexylamine (linear alkylamine family) | ~1.5 nm | Ambient air, $H_2O_2$, organic solvent, bases, $H_2$ annealing (≥250 °C) | Organic acids | Solvothermal treatment | Size of container |
| $AlO_x$ | 2-30 nm | Ambient air, organic solvent | Acids, bases, $H_2$ annealing | Atomic layer deposition | Scaling limited by vacuum, chamber size and processing throughput |
| PMMA | ~100 nm | Ambient air | Organic solvent, $H_2$ annealing | Spin coating | Limited spatial resolution in patterning, too thick for spacer in 2D vertical heterostructures |
| Graphene/hBN | ~4 Å | Ambient air, organic solvent, acids, bases, $H_2$ annealing | - | 2D materials transfer method | Scaling limited by graphene/hBN area and transfer technique, non-removable |
| Parylene | 30-300 nm | Ambient air | - | Thermal evaporation | Scale limited by vacuum, chamber size; processing throughput, non-removable |
| Aryl diazonium | Molecular scale (non specific) | Ambient air | - | Wet chemistry | Non-removable |
| Octadecyltrichlorosilane | Monolayer | Ambient air | - | Wet chemistry | Limited processing throughput due to its toxicity, non-removable |